\documentclass[epj]{svjour}
\usepackage{latexsym}
\usepackage{graphicx}
\usepackage{amssymb}
\usepackage{amsmath}
\usepackage{cite}
\begin{document}
\title{Nuclear processes in Astrophysics: Recent progress}
\author{V. Liccardo\inst{1} \and M. Malheiro\inst{1} \and M. S. Hussein\inst{1,2,3} \and B. V. Carlson\inst{1} \and T. Frederico\inst{1}
}                     
\offprints{}          
\institute{ITA-Instituto Tecnol\'ogico de Aeron\'autica, Pra{\c c}a Marechal Eduardo Gomes, 50 - Vila das Ac\'acias, CEP 12.228-900 S\~ao Jos\'e dos Campos - SP - Brasil \and Instituto de Estudos Avan\c{c}ados, Universidade de S\~{a}o Paulo C. P. 72012, 05.508-970 S\~{a}o Paulo - SP - Brazil \and Instituto de F\'{\i}sica, Universidade de S\~{a}o Paulo, C. P. 66318, 05.314-970 S\~{a}o Paulo - SP - Brazil}
\date{Received: date / Revised version: date}
%
\abstract{
The origin of the elements is a fascinating question that scientists have been trying to answer for the last seven decades. The formation of light elements in the primordial universe and heavier elements in astrophysical sources occurs through nuclear reactions. We can say that nuclear processes are responsible for the production of energy and synthesis of elements in the various astrophysical sites. Thus, nuclear reactions have a determining role in the existence and evolution of several astrophysical environments, from the Sun to the spectacular explosions of supernovae. Nuclear astrophysics attempts to address the most basic and important questions of our existence and future. There are still many issues that are unresolved such as, how stars and our Galaxy have formed and how they evolve, how and where are the heaviest elements made, what is the abundance of nuclei in the universe and what is the nucleosynthesis output of the various production processes and why the amount of lithium-7 observed is less than predicted. 
In this paper, we review our current understanding of the different astrophysical nuclear processes leading to the formation of chemical elements and pay particular attention to the formation of heavy elements occurring during high-energy astrophysical events. Thanks to the recent multi-messenger observation of a binary neutron star merger, which also confirmed production of heavy elements, explosive scenarios such as short gamma-ray bursts and the following kilonovae are now strongly supported as nucleosynthesis sites.
%
} 
\maketitle
\section{Introduction}
\label{sec:intro}
The understanding of the abundance of the periodic table elements is one of the most studied topics of nuclear astrophysics. There are several nuclear reactions leading to their formation. The most common  formation process by which different elements are produced in the universe is the fusion of two nuclei to form a heavier one. This process is able to release energy for elements lighter than $^{56}$Fe, which is the nucleus with the greatest binding energy per nucleon. Beyond this limit, fusion processes require energy from the system in order to occur. To explain the existence of nuclei with $A$ $> $ 60, we need to have a global view of the existence of the many nuclei known to us.

The process by which new atomic nuclei are created from preexisting nuclei is called nucleosynthesis. The initial composition of the universe was established by primordial nucleosynthesis that occurred moments after the Big Bang \cite{Steigman07}. It was then that the H and He isotopes, which today are by far the most abundant species in the universe, formed to become the content of the first stars. The nuclear primordial gas was made of 76\% of H and D, in a much smaller part, 24\% of $^{3}$He and $^{4}$He, with a prevalence of $^{4}$He, and traces of 10 parts per million of $^{7}$Li and $^{6}$Li \cite{Coc04}.
With the formation of stars, heavier nuclei (C, O, Na, Mg, Si) were synthesized through fusion reactions, a process that continues today. 
Some light elements, such as Li, Be and B were formed during a spallation process, in which cosmic rays interacted with C, N, O atoms present in the Interstellar Medium (ISM).

It is currently accepted that stellar nucleosynthesis leads to the formation of heavier elements that astronomers call $metals$. 
Metals can be ejected into the ISM in the later stages of stellar evolution, through mass loss episodes. Star formation from this enriched material, in turn, results in stars with enhanced abundances of metals. This process occurs repeatedly, with the continual recycling of gas, leading to a gradual increase in the metallicity of the ISM with time. Supernovae and compact object mergers are also important to chemical enrichment. They can eject large quantities of enriched material into interstellar space and can themselves generate heavy elements in nucleosynthesis.

The present paper does not concern itself with a general discussion of nucleosynthesis, apart from a few
introductory comments, but rather it attempts to survey the recent developments, both experimental and theoretical, which have been attained in nucleosynthesis and nuclear astrophysics. For a complementary analysis, the more interested reader is referred to some recent reviews on this subject \cite{Jose11, Bertulani16, Coc17}.
This review is organized as follows: in section \ref{sec:bigbang},  Big Bang nucleosynthesis is briefly discussed, with focus on the   problem of the primordial Li abundance. This is followed by an overview of the physics of element production in stars (section \ref{sec:stellar}), paying particular attention to the solar neutrino problem. In the following section \ref{sec:neutron} we discuss  the synthesis of neutron and proton-rich nuclei, showing the state of art of the s-, r- and p-process models. The latest studies and observations of different astrophysical scenarios (compact object mergers, kilonovae) which promise to clarify the origin of the heaviest elements are reviewed in section \ref{sec:NSBH}. Finally, we present the conclusion with several future prospects. A recent review of some of the material covered in this paper can be found in \cite{Descouvemont13}.

\section{Big Bang nucleosynthesis and the cosmological lithium problem}
\label{sec:bigbang}

The Big Bang Nucleosynthesis (BBN) is the phase of cosmic evolution during which it is thought that the primordial nuclei of  light elements, in particular, D, $^{3}$He, $^{4}$He and $^{7}$Li, were formed. 
In 1957 Burbidge $et$ $al$ \cite{Burbidge57} hypothesized reactions that could have arisen within the stars in order to achieve results that agree well with the observations. Actually, the abundances calculated by the four scientists were quite close to those observed. In addition, the observation of the Cosmic Microwave Background (CMB) radiation \cite{Penzias65}, corresponding to a blackbody spectrum with a temperature of 2.73 K, was of paramount importance in this regard. Today we know that the lighter elements (H, D and He) were formed mainly in the moments immediately after the Big Bang, while the elements heavier than He, were actually synthesize in the centers of stars (the first evidence that metals could form in stars was in the early 50s \cite{Merrill52}). After the first initial flare, the universe started to expand and cool down to the point that when it reached the age of 10$^{-6}$ seconds, the temperature had dropped to 10$^{12}$ K. It was then that the primordial nucleons were formed from the quark-gluon plasma. At this moment, quarks combined together in threes to form protons and neutrons. The extremely compressed cosmic soup was now made up of photons, electrons and nucleons. The universe was fully ionized. Photons were absorbed by the atoms then retransmitted immediately. They were trapped in the plasma and no constituents could escape. At this time the universe was completely opaque.
Subsequently, at an age of between three and five minutes from the beginning, the temperature dropped further to reach one billion degrees, and the density was reduced drastically. Under these conditions, the collisions between protons and neutrons  became very effective in forming the first nuclei (D, $^{3}$He, $^{4}$He, $^{7}$Li and $^{7}$Be). 
As the universe continued to expand and cool down very quickly, the temperature dropped enough to allow electrons to bind to nuclei and form the first neutral atoms (Recombination). By the end of this so-called Recombination era, the universe consisted of about 75\% H and 25\% He. It also marked the time at which the universe became transparent as electrons were now bound to nuclei, and photons could travel long distances before being absorbed or diffused (Decoupling).  During the further expansion, small, dense regions of cosmic gas started to collapse under their own gravity, becoming hot enough to trigger nuclear fusion reactions between hydrogen atoms. These were the very first stars to light up the universe. The force of gravity began to pull together huge regions of relatively dense cosmic gas, forming the vast, swirling collections of stars we call galaxies. These in turn started to form clusters, of which one, the so-called Local Group, contains our own Milky Way galaxy.

\begin{figure}
\resizebox{0.48\textwidth}{!}{%
  \includegraphics{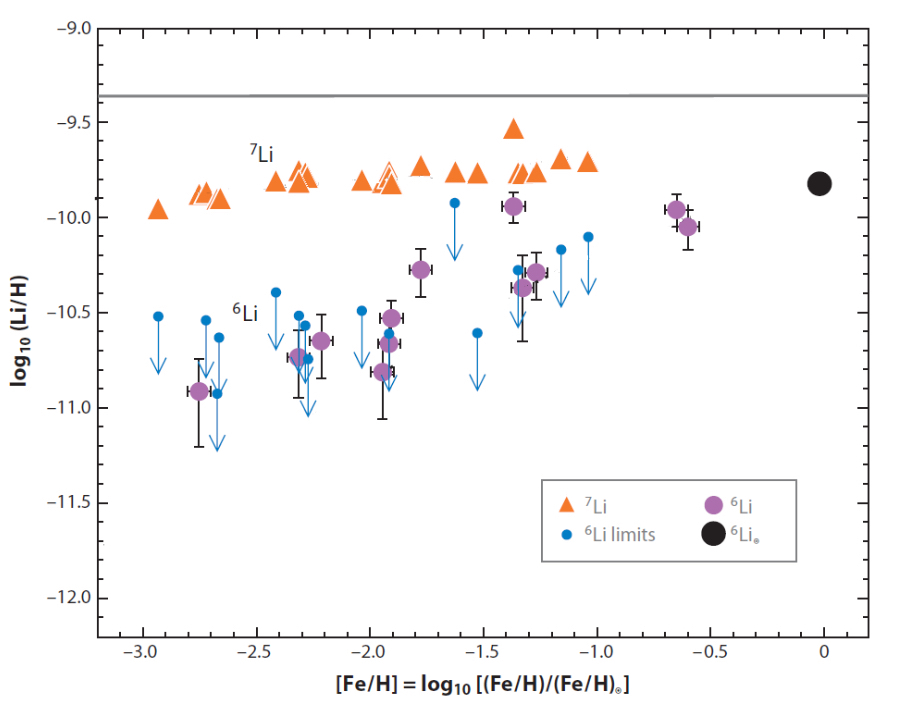}
}
\caption{Lithium abundances in selected metal-poor galactic halo stars. The values obtained for both Li isotopes are shown as a function of metallicity. The metallicity-independence of $^{7}$Li  is known as Spite plateau. It indicates that the abundance of Li is unrelated to galactic nucleosynthesis and therefore represents primordial abundance. The horizontal line is the predicted $^{7}$Li abundance from BBN and the baryon density as determined by WMAP. Data collected from \cite{Fields11}.}
\label{lith7}       
\end{figure}

Within the framework of the standard BBN theory, precise predictions of primordial abundances are feasible since they rely on well-measured cross-sections and a well-measured neutron lifetime \cite{Coc14}. Indeed, even prior to the Wilkinson Microwave Anisotropy Probe (WMAP) era, theoretical predictions for D, $^{3}$He, and $^{4}$He were reasonably accurate \cite{Boesgaard76}. However, uncertainties in nuclear cross-sections leading to $^{7}$Be and $^{7}$Li were relatively large. This led to the problem of the primordial $^{7}$Li abundance, which still represents a challenging issue today \cite{Fields11}. The amount of Li created through the BBN depends primarily on the relative amounts of light and matter.  This is obtained by the photon-to-baryon ratio, which we can measure from the CMB radiation. The Li abundance measured in the galaxy and that predicted by the BBN theory are not consistent.
Indeed, the predicted primordial Li abundance is about a factor of three higher than the abundance determined from absorption lines seen in a population of metal-poor galactic halo stars \cite{Asplund06}. These stars, some of which are as old as their own Galaxy, act as an archive of the production of primordial Li. This conclusion is supported by numerous data on the presence of light elements in the atmospheres of metal-poor stars \cite{Tan09}, in which the interstellar matter was incorporated early in their condensation. Due to convective motion, the surface material of such stars can be dragged into the inner regions, where the temperature is higher and Li is depleted. This effect is evidenced by the low Li abundance in the cold stars of the halo, which are fully convective stars. However, the hottest and most massive stars have only a thin convective layer on the surface, showing no correlation between Li abundance and a temperature \cite{Lyubimkov16}. Figure \ref{lith7} shows the abundance of Li and Fe in a sample of halo stars. The [Li/H] ratio is practically independent of [Fe/H]. Heavy elements are produced by stellar nucleosynthesis and thus their abundance increases over time as the matter circulates in and out of the stars. The independence of Li abundance from that of Fe indicates that $^{7}$Li is not related to galactic nucleosynthesis and therefore is of primordial origin. This flat trend is known as the $Spite$ $Plateau$ \cite{Spite82}, and its value represents the primordial abundance of Li. Li abundance has been measured in several galactic halo poor-metal stars \cite{Fields11, Asplund06}. According to the observations the currently accepted value is [Li/H] = ($1.23_{ - 0.32}^{ + 0.68}$) $\times$ $10^{-10}$, where 95\% of the error is systematic \cite{Ryan00}. In addition, Li was observed in stars of a metal-poor dwarf galaxy and [Li/H] abundances were consistent with the Spite plateau, indicating its universality \cite{Monaco10}. Furthermore, a recent study \cite{Howk12} has shown agreement between the Li abundances of low-metallicity gas in the  ISM of the Small Magellanic Cloud and the value of BBN + WMAP predictions. The measured $^{7}$Li abundance, which is independent of stars, provides an alternative constraint on the primordial abundance and cosmic evolution of Li that is not susceptible to the in situ modifications that may affect stellar atmospheres. However, the results show a disagreement between predictions and measurements if there is any post-BBN Li production, as would be expected.

$^{6}$Li production in primordial nucleosynthesis is much lower than that of the more massive isotope $^{7}$Li. $^{6}$Li and $^{7}$Li are spectroscopically distinguishable by the shift in their atomic lines. This isotopic split is much smaller than the thermal doppler broadening of the Li lines occurring in the stars. Nevertheless, high spectral resolution measurements reach the precision needed to determine the presence of $^{6}$Li. The isotope was observed in several poor-metal halo stars and an isotopic ratio of [$^{6}$Li/$^{7}$Li] $\simeq$ 0.05 was obtained \cite{Asplund06}. Figure \ref{lith7} shows a metallicity-independence pattern also for the [$^{6}$Li/H] abundance, suggesting a primordial origin. $^{6}$Li observations remain controversial: some studies argue that stellar convective motions may have altered the delicate shape of spectral lines thus mimicking the presence of $^{6}$Li \cite{Cayrel07}. A prudent approach is to consider the observations of the abundance $^{6}$Li as an upper limit. In any case, the analysis of this isotope also confirms that most of the primordial lithium is in the form of $^{7}$Li. 

The precise mapping of the primordial anisotropies obtained by WMAP has been used to test the main cosmological models such as the BBN. The CMB radiation carries a record of the conditions in the early universe at a time when H and He nuclei recombined with electrons to form neutral atoms. As a result, photons decoupled from baryons and the universe became transparent to radiation. It is expected that there would be temperature anisotropies in different parts of the present microwave sky, reflecting the oscillations in the photon-baryon plasma around the time of the decoupling (the so-called $baryonic$ $acoustic$ $oscillations$). The measurement of the cosmic baryon-to-photon ratio, $\eta_{WMAP}$ = (6.19 $\pm$ 0.15) $\times$ $10^{-10}$ is one of the most accurate results obtained by WMAP \cite{Komatsu09}. Before the WMAP measurements, $\eta$ was the only free parameter in the BBN model, and the only way to obtain its value was from the observed abundances of D, $^{4}$He, and $^{7}$Li. The new WMAP baryon density is much more accurate and allows us to eliminate the last free parameter in the BBN theory, providing a new way for verifying the validity of this model. Using $\eta_{WMAP}$ as input in the BBN model and propagating errors, it is possible to compare the expected and  observed abundances for all light elements. The observations of D and $^{4}$He are in agreement with the theory (the abundance measured at z $\sim$ 3, the theoretical predictions at z $\sim$ $10^{10}$ and the WMAP data at z $\sim$ 1 are all consistent). [$^{3}$He/H] measurements have not been reported since they are still unreliable. As for Li, the BBN + WMAP expectation and the measurements are in complete disagreement: using $\eta$ = $\eta_{WMAP}$ the expected abundance of Li becomes [Li/H]$_{BBN+WMAP}$ $\simeq$ (5.1 $\pm$ 0.6) $\times$ $10^{-10}$, quite different from that given by the Spite plateau. In conclusion, the theoretical value is above the observations by a factor $\simeq$ 2.4-4.3, which represents a discrepancy of 4.2-5.3 $\sigma$.

A number of explanations were put forward to solve the cosmological problem of Li. The issues to consider for a possible solution are numerous and belong to different fields of physics. They included (i) astrophysical solutions: the problem may lie in the observations, leading to an incorrect estimation of the primordial Li abundance, (ii) nuclear solutions:  reevaluate rate and cross-sections reactions that lead to the creation or depletion of $A$ = 7 nuclei, (iii) solutions beyond standard physics: by building new models that go beyond the standard models of cosmology and particle physics. Regarding an astrophysical solution, the Li problem might be hidden behind an underestimation of the Spite plateau, caused by systematic errors or improper determination of stellar temperatures. In fact, in the relevant stars, Li is mostly ionized. The measured abundance (obtained from the neutral Li line at 670.8 nm) thus needs a correction based on the [Li$^{+}$/Li$^{0}$] ratio, which varies exponentially with temperature. Another point to consider is to whether the current Li content in the stars is actually primordial or whether factors that caused its depletion have occurred over time such as convective motions, turbulence or gravitational effects. More radical solutions involve elements that go beyond the standard models of cosmology and/or particles and could point to new physics. Some observations might bring into question the cosmological principle, highlighting large-scale inhomogeneities. If the baryon-to-photon ratio varied depending on the inhomogeneity, the BBN could have been different \cite{Fields11}.

According to the BBN nuclear network, about 90\% of Li is produced by Be decay, while only 10\% is produced directly in primordial nucleosynthesis. $^{7}$Be is radioactive and decays, with a half-life of about 53 days, into stable atoms of $^{7}$Li through the only energy-efficient decay channel: electronic capture. This process takes place at later stages in BBN, since the $^{7}$Be electronic capture probability is virtually zero in the primordial universe due to the low electronic density \cite{Khatri11}. Actual abundance of $^{7}$Li is directly related to the quantity of $^{7}$Be produced in the BBN, even though it is not possible to observe $^{7}$Be in the stars at present, as it has now completely decayed. An investigation of the processes that created and destroyed $^{7}$Be through the history of the universe could resolve the disagreement between predictions and observations of Li. The problem could, in fact, be caused by a poor estimate of the production and destruction rates of Be, probably due to an incorrect calculation of cross-sections. The main Be production reaction, $^{3}$He($\alpha$,$\gamma$)$^{7}$Be, has been well studied and the cross-section is known with a precision of approximately 3\% \cite{Confortola07, deBoer14}. Destruction of Be occurs mainly through the reactions $^{7}$Be(n,p)$^{7}$Li and $^{7}$Be(n,$\alpha$)$^{4}$He. Data on these reactions are scarce or completely missing, thus affecting the prediction abundance of $^{7}$Li  by the BBN. Regarding the $^{7}$Be(n,p)$^{7}$Li reaction no cross-section measurements have yet been made in the MeV region, i.e., in the energy range of interest for the BBN. A measure of (n,p) at higher energies is scheduled at the Time-of-Flight facility (n\_TOF) at CERN \cite{Barbagallo16}. A secondary contribution is given by the $^{7}$Be(n,$\alpha$)$^{4}$He process which accounts for only 2.5\% of the total Be destruction, with an associated uncertainty of 10\%. It has never been tested in the range of BBN temperatures and only one measurement, at thermal neutron energy, has been performed so far by Bassi $et$ $al$ \cite{Bassi63} at the ISPRA reactor. Several theoretical extrapolations have been carried out, yielding various estimated cross-section trends with discrepancies of up to two orders of magnitude \cite{Hou15}. A newly measured S-factor for the T($^3$He,$\gamma$)$^6$Li reaction rules out an anomalously-high $^6$Li production during the Big Bang as an explanation to the high observed values in metal-poor first generation stars \cite{Zylstra16}. Nevertheless, the energy-dependent $^{7}$Be(n,$\alpha$)$^{4}$He cross-section, has been recently measured for the first time from 10 meV to 10 keV neutron energy \cite{Barbagallo16(2)}. Results are consistent, at thermal neutron energy, with the only previous measurement performed in the 60s at a nuclear reactor, but regarding the trend of the cross-section as a function of neutron energy, the experimental data are clearly incompatible with the theoretical estimate used in BBN calculations. 

\begin{figure}
\resizebox{0.48\textwidth}{!}{%
  \includegraphics{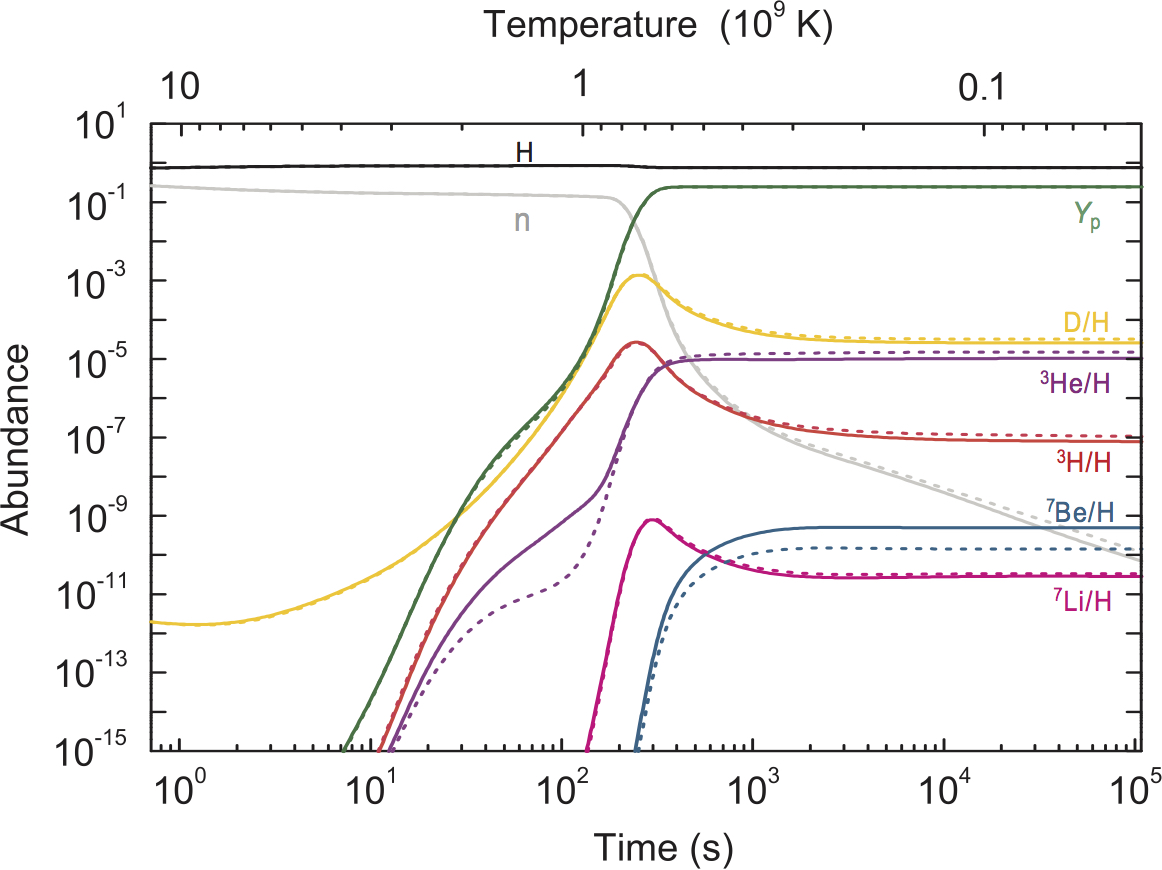}
}
\caption{Distribution of early primordial light elements in the universe as a function of time and temperature. The model (dotted lines) successfully predicts a lower abundance of the Be isotope - which eventually decays into Li - relative to the classical Maxwell-Boltzmann distribution (solid lines), without changing the predicted abundances of D or He \cite{Hou17}.}
\label{extensive}       
\end{figure}

Furthermore, in a recent publication by Hou $et$ $al$ \cite{Hou17} a team of scientists has proposed an elegant solution to the problem. One assumption in the BBN is that all the nuclei involved in the process are in thermodynamic equilibrium and their velocities follow the classical Maxwell-Boltzmann distribution. The authors claim that Li nuclei do not obey this classical distribution in the extremely complex, fast-expanding Big Bang hot plasma. By applying what is known as non-extensive statistics \cite{Bertulani13}, the problem can be solved instead. The modified velocity distribution found with these statistics violates the classical distribution by a very small deviation of about 6.3-8.2\%, and can successfully and simultaneously predict the observed primordial abundances of D, He, and Li (Fig. \ref{extensive}).

Alternative proposed solutions to the problem are the modification of the expansion rate during BBN \cite{Coc06} or depletion of $^{7}$Li in stellar interiors \cite{Michaud1991}. Clearly, the problem of $^{7}$Li is still an open issue in astrophysics and it represents an exciting challenge for the future.

\section{Stellar nucleosynthesis}
\label{sec:stellar}

The nucleosynthesis of elements heavier than those produced in the BBN requires the extreme temperatures and pressures found within stars. Charged-particle reactions are the main source of energy generation in stars and are responsible for the creation of elements up to Fe \cite{Iliadis07}.
It is beyond the scope of this review to describe the highly complex phenomena of nucleosynthesis. The interested reader is referred to \cite{Karakas14} for a review of stellar nucleosynthesis. 
In this section, we will focus mostly on nucleosynthesis in massive stars ($>$ 8 $M_{\odot}$), since this is the critical mass for a star necessary to start the fusion of C after the central He density has been exhausted. After three further fusion phases, an iron core is formed and no further energy can be gained. These processes began as H and He from the Big Bang collapsed into the first protostars at 250 million years \cite{Iben65}. Star formation has occurred continuously in the universe since that time. Once a critical ignition temperature of $\sim$ 10$^{6}$ K is achieved, protostars begin to fuse D. The start of central H-burning is the start of the Main Sequence (MS) phase. Then the conversion of central H into He ($T$ $\sim$ 10$^{7}$ K) occurs with energy production and with the simultaneous destruction (due to the high temperatures) of the newly synthesized Li, Be and B nuclei. Nevertheless, these atoms are reformed, although in small quantities, in the ISM by the action of cosmic rays \cite{Vangioni00}. The release of nuclear energy by fusion prevents further contraction of the star. The needed energy is released by the nuclear reactions of the carbon-nitrogen-oxygen cycles ($CNO$-cycles) and the proton-proton chains (pp-chains). Both, the pp-chain and the $CNO$-cycles, produce a He nucleus from four H nuclei \cite{Salpeter52, Bethe39}. The central energy release establishes a temperature and pressure gradient, falling from the center to the surface, creating a force acting in the opposite direction to gravity and keeping the star stable. During the MS massive stars have an inner convective core and a radiative envelope, i.e., in the core the energy is mainly transported outwards by the flow of matter, while in the envelope by it is transported by the outward diffusion of photons. Stars, in general, spend about 90\% of their lifetime on the MS with a rather constant surface temperature and luminosity.

The end of the MS phase is marked by the exhaustion of H in the stellar center. The missing energy production leads to a short contraction phase of the whole star, until the H-burning ignites in a shell around the core, leading to an expansion of the envelope, while the core still contracts until He-burning starts. The core reaches a temperature of $\approx$ 1.5 $\times$ $10^{8}$ K, and it is at this time that the creation of C nuclei begins  \cite{Hoyle46}. A nucleus of $^{12}$C is formed through a reaction involving three $^{4}$He nuclei, the so-called 3-$\alpha$ $process$ \cite{Salpeter52a}. The process is favored by both the high temperature in the core of stars, which allows the electrostatic repulsion to be overcome, and by the high $^{4}$He density. With the formation of $^{12}$C via the 3-$\alpha$ process, subsequent $\alpha$-capture reactions can occur to produce $^{16}$O. The 3-$\alpha$ process has been revisited in many theoretical studies in recent years, with considerably discrepancies seen between the works (e.g., \cite{Angulo99, Siess02, Fynbo05, Yabana12}). 
Using new theoretical approaches, a very large reaction rate at low temperatures was obtained by Ogata $et$ $al$ \cite{Ogata09} resulting in a significant impact on the astrophysics and producing results that are incompatible with observations \cite{Suda11, Matsuo11, Peng10, Saruwatari10}. Subsequent works, with an improved theory, did not find the large enhancement \cite{Nguyen12, Nguyen13}, and have successfully produced the 3-$\alpha$ reaction rate in the low temperature regime, where measurements are impossible without extrapolation. 
After H and He core burning, low and intermediate mass stars (0.8-8 $M_{\odot}$) develop an electron-degenerate C-O core which is too cool to ignite C-burning. The convective envelope moves down and penetrates the He layer until it reaches the boundary of the core. At this point in time, a double shell structure develops: the center of the star is formed by a contracting degenerate C-O core, which is
surrounded by both a He and an H-burning shell. This structure is unstable and He shell flashes develop, resulting in thermal
pulses that are characteristic of Asymptotic Giant Branch (AGB) stars \cite{Herwig05}. During this Thermally Pulsing (TP-AGB) phase, the star has a high mass loss rate, which removes the entire envelope in a short time, leaving a naked C-O core which evolves into a White Dwarf (WD) (see \cite{Koester90} for a review). 

In massive stars, on the contrary, the further increase of temperature due to the contraction and ignition of the C-O core causes the fusion of $^{12}$C atoms through which  $^{20}$Ne, $^{23}$Na, and $^{24}$Mg are created. Massive stars will proceed to synthesize by $\alpha$-capture Ne to O and Mg during a Ne-burning phase; O and Mg to Si and S during an O-burning phase and finally in Si-burning, Si and S are built into iron-group elements like Ni, Fe, and Cr. These burning phases ignite in this order with increasing temperatures, from $T$ $\geqslant$ 6 $\times$ $10^{8}$ K for C-burning to $T$ $\geqslant$ 3.5 $\times$ $10^{9}$ K for Si-burning. With increasing ignition temperatures, the time scales of the burning phases also decrease, so while C-burning is of the order of 1000 years, Si is exhausted in the center in a couple of days. The shorter time scales of the later burning phases are related to the increasing energy loss by neutrinos with increasing temperatures. Because the maximum of nuclear binding energy occurs near Fe, stars cannot gain energy by fusion after Si-burning. As a consequence, the iron core grows until it reaches the critical mass, the Chandrasekhar mass ($M_{Ch}$ $\approx$ 1.38 $M_{\odot}$), and then collapses, marking the death of the star. Prior to collapse, a massive star will eventually consist of concentric onion-like layers burning different elements. Lighter elements will be produced in the outer layers, moving progressively through the $\alpha$-$ladder$ towards the interior of the star, with Fe and Ni at the core. 
During the final moments, electron degeneracy pressure in the core will be unable to support its weight against the force of gravity. As a result, the collapse increases temperatures in the core, releasing very high-energy gamma rays. These high-energy photons break the Fe nuclei up into He nuclei through a photodisintegration process. At this stage, the core has already contracted beyond the point of electron degeneracy, and as it continues contracting, protons and electrons are forced to combine to form neutrons (Proto-Neutron Star (PNS)). This process releases vast quantities of neutrinos carrying substantial amounts of energy, again causing the core to cool and contract even further. The contraction is finally halted by the repulsive component of the nuclear force once the density of the inner core is nearly twice that of the atomic nucleus. The abrupt halt of the collapse of the inner core and its rebound generates an outgoing shock wave which reverses the infalling motion of the material in the star and accelerates it outwards. Such explosions are called Core-Collapse Supernovae (ccSNe) (see Section \ref{sec:Supernovae}). The result of this dramatic explosion is the birth of a Neutron Star (NS) or a Black Hole (BH) (see, for more details, \cite{Zwicky39, Baade34}). The products of stellar nucleosynthesis are dispersed into the ISM through mass loss episodes (planetary nebulae), the stellar winds of low-mass stars and ccSNe. ccSNe not only serve as the mechanism for the creation of heavier elements, they also serve as the mechanism for their dispersal.

Stellar reaction rates are frequently uncertain since they are often based on theoretical extrapolations of experimental data taken at energies that are substantially higher than the ones in stellar interiors. The reaction rates define the timescale for the stellar evolution, dictate the energy production rate, and determine the abundance distribution of seed and fuel for the next burning stage. The extrapolations are based on theoretical assumptions such as the incompressibility of nuclei, reducing the probability of fusion \cite{Jiang11}, and the possible impact of molecular cluster formation during the fusion process, which increases the fusion probability \cite{Goasduff14}. This uncertainty range may have severe consequences for our understanding of stellar heavy ion burning associated with late stellar evolution and stellar explosions \cite{Gasques07}. New measurements of the $^{12}$C+$^{16}$O fusion reaction need to be performed over a wide energy range to provide a more reliable extrapolation of the fusion cross-section and to reduce substantially the uncertainty for stellar model simulations. Some other poorly known nuclear reactions that are of particular importance for stellar evolution are capture and fusion reactions (e.g., $^{3}$He($\alpha$,$\gamma$)$^{7}$Be, $^{14}$N(p,$\gamma$)$^{15}$O, $^{17}$O(p,$\gamma$)$^{18}$F) and heavy-ion reactions (e.g., $^{12}$C+$^{12}$C, $^{16}$O+$^{16}$O) which influence the subsequent evolutionary stages of massive stars \cite{Arcones17}. Recently, advances have been made in this context. More than a thousand stellar models aimed at exploring properties of pre-supernova massive stars and C-O WDs have been proposed by Fields $et$ $al$ \cite{Fields16, Fields18}. The group found that experimental uncertainties in reaction rates (e.g., 3-$\alpha$, $^{14}$N(p,$\gamma$)$^{15}$O, $^{12}$C($\alpha$,$\gamma$)$^{16}$O,  $^{12}$C($^{12}$C,p)$^{23}$Na,  $^{16}$O($^{16}$O,n)$^{31}$S, $^{16}$O($^{16}$O,$\alpha$)$^{28}$Si) dominate the variations of properties of both the progenitor and the CO core, such as burning lifetime, composition, central density, the core mass and O-depletion. The study allowed identification of the reaction rates that have the largest impact on the variations of the properties investigated and suggests that the variation in properties of the stellar model grows with each passing phase of the evolution towards Fe core-collapse.
More progress is required to significantly improve the accuracy of stellar models and to provide more reliable nucleosynthesis predictions for nuclear astrophysics. The current generation of stellar models is still affected by several, not negligible uncertainties related to our poor knowledge of thermodynamic  processes and nuclear reaction rates, as well as the efficiency of mixing processes. These drawbacks have to be properly taken into account when comparing theory with observations to obtain evolutionary properties of both resolved and unresolved stellar populations. 

The synthesis of heavy elements in massive stars provides an important source for the chemical enrichment of their surrounding ISM and hence of the universe. The ejecta of stars, the stellar yields, have been calculated by various groups since the early stage of stellar evolution modelling \cite{Gibson02}. Each stellar mass can produce and eject different chemical elements and the yields are therefore a function of the stellar mass but also of the original stellar composition. Low and intermediate mass stars  produce He, N, C and heavy s-process elements (Section \ref{sec:AGB}). Stars with $M$ $<$ 0.8 $M_{\odot}$  do not contribute to the galactic chemical enrichment and have lifetimes longer than the Hubble time. Massive stars ($M$ $>$ 8-10 $M_{\odot}$ ) produce mainly $\alpha$-elements (O, Ne, Mg, S, Si, Ca), some Fe, light s-process elements, perhaps r- and p-process elements and explode as ccSNe. 
Predictions from Romano $et$ $al$ \cite{Romano10} concerning the abundance of several chemical elements, obtained by using different sets of stellar yields and compared to observations in stars, are in agreement for some chemical species, whereas for others the agreement is still very poor. The reason for this resides in the uncertainties still existing in the theoretical stellar yields. {C{\^o}t{\'e} $et$ $al$ \cite{Cote16} have compiled several observational studies to constrain the values and uncertainties of fundamental input parameters, including the stellar initial mass function and the rate of supernova explosions. The uncertainties they found, which are lower limits, depend on the galactic age and on the targeted elemental ratio. Portinari $et$ $al$ \cite{Portinari98} and Marigo \cite{Marigo01} have published the most recent complete self-consistent sets of yields from AGB and massive stars. Drawbacks of this set are that AGB models are not based on full stellar evolution simulations (synthetic models), the explosion yields are not consistent (based on yields by Woosley and Weaver \cite{Woosley95}) and they include only isotopes up to Fe. Because of the lack of alternatives, these yields are still in use (e.g., \cite{Vogelsberger13, Yates13}). 
Despite several considerable improvements in the field of stellar nucleosynthesis in recent years, no single combination of stellar yields is found which is able to reproduce at once all the available measurements of chemical pattern and abundance ratios in the Milky Way. Further efforts are needed to improve and expand the current existing grids of stellar yields and to better understand their dependence on the metallicity, mass, and rate of rotation. A further discussion on stellar nucleosynthesis is given in Section \ref{sec:AGB}.

\subsection{Solar neutrino physics: current status}

The production of energy in the stars through nuclear reactions was demonstrated only in the 60s. Thanks to the predicted solar neutrino flux \cite{Bahcall64} and to the following detection \cite{Davis64} of these neutrinos emitted during the H-burning it was possible to confirm the presence of thermonuclear reactions in the interior of the Sun. Neutrinos created by H-burning can escape almost without interaction straight from the heart of the star. In so doing, they carry information on what is actually happening at the center. H-burning necessarily involves the emission of neutrinos. They arise when the nuclear weak interaction changes a proton to a neutron $p$ $\rightarrow$ $n$ + $e^{+}$ + $\nu_{e}$. This must occur twice during the H-burning process $4p$ $\rightarrow$ $^{4}$He + $2e^{+}$ + $2\nu_{e}$. The expected flux of neutrinos can be found by noting that the formation of each $^{4}$He is accompanied by the release of two neutrinos and a thermal energy $Q_{eff}$ of about 26 MeV. Results obtained from solar neutrino detection have been explained with theoretical predictions that account for neutrino flavor oscillations. It is interesting to point out that this explanation was proposed more than thirty years ago and the problem is discussed here from a historical perspective. The following sections outline what is left to do in solar neutrino research and solar abundances models. We also discuss the main results from recent lines of research.

The solar neutrino problem was definitively solved by combining the SNO and the Super-Kamiokande measurements in 2001 \cite{Ahmad01}. The SNO collaboration determined the total number of solar neutrinos of all types (electron, muon, and tau) as well as the number of electron neutrinos alone. The total number of neutrinos of all types agrees with the number predicted by the solar model. Electron neutrinos constitute about a third of the total number of neutrinos. The missing neutrinos were actually present but in the form of the more difficult to detect muon and tau neutrinos.
The agreement between theoretical predictions and observations was achieved thanks to a new understanding of neutrino physics, which required a modification of the Standard Model to permit neutrinos oscillations \cite{Ahmad02}. There are still open questions that remain to be addressed, including what is the total solar neutrino luminosity, what is the $CNO$ neutrino flux of the Sun and what is the solar core metallicity.

The latest predictions of the Standard Solar Model (SSM) and the neutrino fluxes from various processes taking place inside the Sun are given in Vinyoles $et$ $al$ \cite{Vinyoles17} and Bergstr{\"o}m $et$ $al$ \cite{Bergstr16}. As expected, the neutrinos originate from the primary reaction of the proton-proton chain, the so-called $pp$ neutrinos, and constitute nearly the entirety of the solar neutrino flux, vastly outnumbering those emitted in the reactions that follow. These neutrinos have low energy, never exceeding 0.42 MeV. The neutrinos from electron capture by $^{7}$Be (the reaction which initiates the ppII reaction branch) are the next most plentiful. The $^3$He($\alpha$,$\gamma$)$^7$Be reaction is the starting point of the ppII and ppIII chains  in the solar H-burning, therefore its rate has a substantial impact on the solar $^7$Be and $^8$B neutrino production. Using the SSM, the flux of these neutrinos can be calculated \cite{Serenelli13}. The reaction $^3$He($\alpha$,$\gamma$)$^7$Be is one of the most uncertain, even if many experiments have been done in the last decade, clearing up several long-standing issues \cite{Adelberger11}. Most of these cross-section measurements are concentrated in the low energy range and an higher level of precision is needed. Furthermore, there is no experimental data above 3.1 MeV, and there are conflicting data sets for the $^6$Li(p,$\gamma$)$^7$Be reaction that have an impact on the level scheme of $^7$Be \cite{He13}.

In the last years, much effort has been applied to the understanding of solar neutrino physics, including nuclear physics experiments, improvements of the SSM model and extensive solar neutrino measurements at new facilities such as Super-Kamiokande \cite{Abe14, Haubold14} and Borexino \cite{Bellini10}. Plans at Borexino \cite{Bellini2016} include improvements in purity to reduce backgrounds and enable the first detection of neutrinos from the $CNO$ cycle \cite{Davini16}. SNO+ \cite{Asahi15} is nearing completion and promises detection of proton-electron-proton (pep) as well as $CNO$ neutrinos. The pep reaction produces sharp-energy-line neutrinos of 1.44 MeV. Detection of solar neutrinos from this reaction was reported by  Bellini $et$ $al$ \cite{Bellini12} on behalf of the Borexino collaboration.
The detection of $CNO$ neutrinos, together with further constraints from precision measurements of the $^{8}$B neutrino flux, offer the most promising pathway to determine the metal content of the Sun, which is still under debate (see \cite{Delahaye06, Bahcall06, Asplund09, Bergemann14} and references therein). Neutrinos produced by $^{8}$B  $\rightarrow$ $^{8}$Be + $e^{+}$ + $\nu_{e}$ have very small flux, about 4.9 $\times$ $10^{6}$ cm$^{-2}$ s$^{-1}$, but because of their high-energy (up to $\sim$ 15 MeV) the $^{8}$B neutrinos dominate the fluxes of many chlorine and water detectors, such as the more recent SNO and Super-Kamiokande experiments. Despite the improvements obtained in the last measurements, the flux uncertainty associated with these neutrinos ($\pm$10\%) is still rather large \cite{Furuno11}. The uncertainty for the $^{7}$Be neutrinos is smaller than that for $^{8}$B neutrinos, about $\pm$5\% \cite{Bellini11} and the flux of pp-neutrinos is accurately determined to the level of 1\%. Another factor to be taken into account is the different temperature-dependence of each reaction. The $^{8}$B neutrinos depend on temperature as $T_c^{25}$, where $T_{c}$ is the characteristic one-zone central temperature \cite{Bahcall96}. The $^{7}$Be neutrinos vary as $T_c^{11}$. The pp-neutrinos are relatively less dependent on the central temperature. It is, however, worth mentioning the efforts made to accurately determine the cross-section of the $^7$Be(p,$\gamma$)$^8$B reaction. The $^7$Be(p,$\gamma$)$^8$B  cross-section directly affects the detected flux of $^8$B decay neutrinos from the Sun and plays a crucial role in constraining the properties of neutrino oscillations \cite{Haxton13}. The $^7$Be(p,$\gamma$)$^8$B reaction also plays an important role in the evolution of the first stars which formed at the end of the cosmic dark ages \cite{Bromm04}. The precision of the astrophysical S-factor at solar energies ($\sim$ 20 keV) is limited by extrapolation from laboratory energies of typically 0.1-0.5 MeV \cite{Adelberger11}. The theoretical predictions have uncertainties of the order of 20\% \cite{Bahcall04}, whereas recent experiments were able to determine the neutrino flux emitted from $^8$B decay with a precision of 9\% \cite{Ahmed04}. Direct measurements were carried out with a radioactive $^7$Be beam on an H$_{2}$ target \cite{Gialanella00}, or with a proton beam on a $^7$Be target \cite{Baby03}. Furthermore, indirect measurements were performed through Coulomb dissociation \cite{Davids01, Schumann03, Schumann06} as well as the transfer reaction method \cite{Ogata03, Das06}. More recently the astrophysical S-factors and reaction rates of $^7$Be(p,$\gamma$)$^8$B have also been  investigated by Du $et$ $al$ \cite{Du15}.

In addition to advanced neutrino detection experiments, titanic efforts are being invested in studying the solar abundance problem. Recent more accurate analyses \cite{Asplund09, Caffau11} have indicated that the solar photospheric metallicity is significantly lower than older values \cite{Grevesse98}. Despite these improvements, there is a persistent conflict with standard models of the solar interior according to helioseismology, a discrepancy that has yet to find a satisfactory resolution \cite{Vagnozzi17}. This solar modelling problem has been the subject of a large number of investigations and reviews \cite{Bahcall05, Bahcall06, Delahaye06, Bergemann14}. Many modifications of the solar model have been proposed, including enhanced diffusion\cite{Guzik05}, the accretion model \cite{Guzik10, Serenelli11}, etc. However, none of these modifications has succeeded in solving the problem.
At present, the main task of solar nuclear physics is to improve the precision of the underlying nuclear reaction rates that connect neutrino observations with solar and neutrino physics, and to produce more accurate and realistic stellar models.

\section{Heavy elements nucleosynthesis}
\label{sec:neutron}

Elements much heavier than Fe, such as Pb, Au, U are not produced in standard stellar nucleosynthesis. Their formation involves different processes occurring inside stars or during explosive and catastrophic events. During these events, the capture of neutrons or protons by atoms is the main process by which heavy elements are formed.
 
There are three main processes by which nucleosynthesis of heavier elements occur: the s-, the r- and the p-process \cite{KAPPELER99, Seeger65, Arnould76}.

\begin{itemize}
  \item The s-process (slow) occurs inside massive stars (weak s-process) and stars that from the evolutionary point of view, are in the AGB phase (main s-process) \cite{Busso99, Bisterzo11, Karakas12}. These stars are able to build heavy and stable nuclei from Fe up to $^{209}$Bi (Fig. \ref{snucl}). The process occurs when a nucleus is able to capture neutrons one at a time. The resulting nucleus can be stable, or if radioactive, decays ($\beta$) into a stable element (following a path that leads to the stability valley) before the next neutron is captured. 
The flow of neutron density in the s-process is estimated to be between 10$^{6}$ and 10$^{11}$ neutrons/cm$^{2}$/s. It is responsible for about half of the heavier isotopes of iron. 
The relevant properties necessary for describing the s-process chain are the neutron capture cross-sections and, in addition, the beta decay rates of these unstable isotopes, which are sufficient long-lived to allow neutron captures to compete.
A well-known element of the s-process is $^{43}$Tc, an element with no stable isotopes that has a half-life of millions of years and was used by Merrill \cite{Merrill52} to prove evidence of nucleosynthesis in other stars, such as S-Type.

 \item The second half of the solar abundances above Fe is contributed by the r-process (rapid) which in contrast to the s-process, occurs at high temperature and neutron density ($N_{n}$ $\gtrsim$ 10$^{20}$ cm$^{-3}$). With these physical conditions a nucleus can absorb a neutron before it $\beta$ decays. Explosive scenarios such as ccSNe, NS mergers, and collapsars are potential r-process sites \cite{Burbidge57, Woosley94, Fujimoto07, Komiya2016}. 
The total process is extremely fast and takes place in a time interval comprising between 0.01 and 10 seconds. This process produces neutron-rich nuclei which then decay gradually towards the valley of stability. This implies that the reaction path is shifted into the neutron-rich region of the nuclide chart until a $waiting$ $point$ is reached. At waiting points, the (n,$\gamma$) sequence is halted by the inverse ($\gamma$,n) reactions due to the hot photon bath of the explosion. Contrary to the s-process, where the abundances are dependent on the cross-section values, the r-abundances are determined by the $\beta$-decay half-lives of these waiting points close to the neutron drip line. More than 90\% of elements such as Eu, Au, and Pt in the solar system are believed to be synthesized in the r-process \cite{Burris00}. 
 
Due to the r-process, it is possible to overcome the $\alpha$-decay barrier of $^{210}$Bi that stops the s-process. For nuclei heavier than $^{210}$Bi, r-process neutron capture is interrupted when it meets the fission threshold induced by neutron absorption. The r-process is the only responsible for the existence of nuclei heavier than Bi, especially long-lived nuclei $^{232}$Th, $^{325}$U, and $^{328}$U.

 \item Some proton-rich nuclides found in nature (mass numbers of $A$ $>$ 74, between $^{74}$Se and $^{196}$Hg) are not reached in the previous processes and therefore at least one additional process is required to synthesize them.  
P-nuclei could well be synthesized by successively adding protons to a nuclide ((p,$\gamma$) captures) or from the destruction of pre-existing s- or r-nuclides by different combinations of $\gamma$-$process$ (($\gamma$,n) reactions followed by ($\gamma$,p) and/or ($\gamma$,$\alpha$) reactions). Some $\beta$-decays, electron captures or (n,$\gamma$) reactions eventually complete the nuclear chain. These reactions may lead directly to the production of a p-nuclide. The resulting p-abundances are influenced by the proton and photon density and by the initial seed abundances. 
The possible astrophysical sites of origin of p-process nuclei in the universe are not well constrained. Neutrino winds from the formation of a NS ($\nu$p-process \cite{Frohlich06}), ccSNe and Type Ia supernovae ($\gamma$-process) have been considered as a possible site for the p-process. It is also conceivable that there is not just a single process responsible for all p-nuclei \cite{Rauscher13}.

\end{itemize}

In the following sections, we will address the main aspects associated with neutron and proton capture nucleosynthesis.
This is supplemented with an overview of observations from explosive processes that characterize ccSNe.

\begin{figure}
\resizebox{0.48\textwidth}{!}{%
  \includegraphics{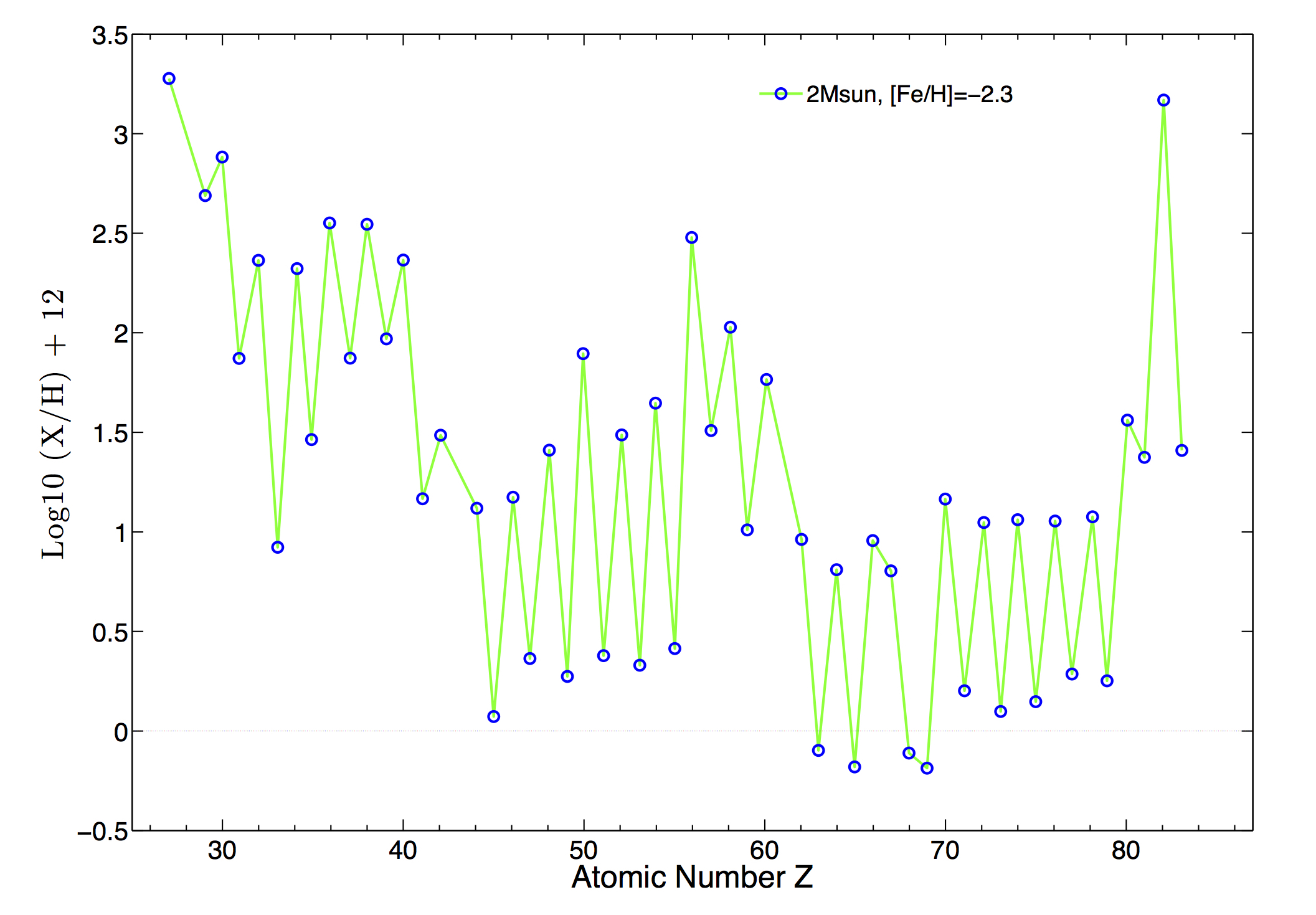}
}
\caption{The s-process element abundance pattern of a 2 $M_{\odot}$ model AGB star with [Fe/H] = -2.3. The s-process operates efficiently enough in this metal-poor star to produce s-process abundances equal to or greater than the solar system values. The s-abundances are from  \cite{Lugaro12}.}
\label{snucl}       
\end{figure}

\subsection{Massive and AGB stars}
\label{sec:AGB}
\subsubsection{s-, and i-processes}

\begin{figure*}
\begin{center}
\resizebox{0.8\textwidth}{!}{%
  \includegraphics{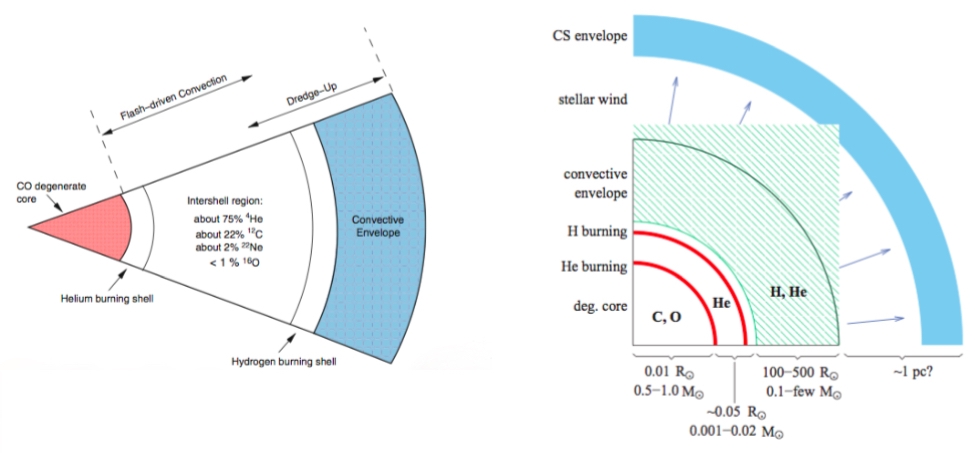}
}
\caption{Schematic structure of a TP-AGB star. The degenerate C-O core is surrounded by the H and He-burning shells. The C-O core is very compact and it is surrounded by a first shell in which the He combustion takes place. Over this layer, there is a region, called the $intershell$ in which the H fusion takes place and rich in He produced from the second shell. Other elements present in non-negligible quantities are C, Ne, and O. As can be seen in the figure, the two shells are very close each other and this is precisely one of the characteristics that induce the thermal pulse and the consequent instability.}
\label{agbstar}
\end{center}       
\end{figure*}

The weak s-process takes place at the end of the convective He-burning in the core and in the subsequent convective C-burning shell \cite{Raiteri91} in massive stars (e.g., 25 $M_{\odot}$), whereas the main s-process occurs in low-mass AGB stars. It is characterized by comparably low neutron densities, so that neutron capture times are much slower than most $\beta$-decay times. This implies that the reaction path of the s-process follows the stability valley. Although the available cross-sections under stellar conditions were very scarce and rather uncertain at the time it was already inferred \cite{Burbidge57} that the product of cross-section times the resulting s-abundance is roughly constant with jumps at shell closures and gaps at branchings.

Weak s-process nucleosynthesis is responsible for the production of the low-mass range of the s-process elements from iron group seed nuclei to $^{58}$Fe on up to Sr and Y (see \cite{Kappeler11} and references therein). The neutron source is provided by the reaction $^{22}$Ne($\alpha$,n)$^{25}$Mg. First attempts to investigate the possible role of rotational mixing on the s-process production in massive stars have shown that this classic picture could be significantly revised \cite{Herwig03}. The impact of rotation on the nucleosynthesis in low-Z massive rotating stars has been explored by different groups \cite{Meynet06, Hirschi07, Pignatari08}. At solar metallicity, rotation-induced mixing has a moderate effect on s-process production. At very low metallicities, however, rotation-induced mixing has a much stronger effect, and therefore a large impact on the evolution and nucleosynthesis of the first stellar generations in the universe. Indeed, rotation leads to mixing between the He-burning core and the H-burning shell. Eventually, the He-burning products $^{12}$C and $^{16}$O are mixed into the H-burning shell, which produces $^{14}$N via the $CNO$ cycle. Later on, the $^{14}$N is mixed back into the He-burning core, at which point it immediately converts into $^{22}$Ne via $^{14}$N($\alpha$,n)$^{18}$F($e^{+}$,$\nu_{e}$)$^{18}$O($\alpha$,$\gamma$)$^{22}$Ne. At the end of He-burning in the core, $^{22}$Ne($\alpha$,n)$^{25}$Mg releases large amounts of neutrons and drastically changes the s-process production \cite{Frischknecht12}. Due to a high neutron number density, the weak s-process possibly produces heavy nuclei up to A $\sim$ 200. Recent observations \cite{Chiappini11, Barbuy14} have confirmed that the s-abundance in globular clusters in the bulge of our Galaxy is compatible with the s-process production in fast-rotating massive stars at low metallicity, supporting the view that massive stars could indeed also be important sources for these elements.
However, studies about the uncertainty of nuclear physics on the s-process \cite{Iliadis15, Cescutti17, Nishimura18} should be taken into account, and production of heavy s-process nuclei in massive low-Z stars strongly depends on rotation. The abundance is also influenced by the uncertainty of neutron source or neutron poison reactions (see \cite{Nishimura14} and references therein). The production of Sr and Ba in metal-poor stars has been investigated because of their observational importance \cite{Hansen14}. Cescutti $et$ $al$ \cite{Cescutti15} presented galactic chemical evolution models using a large grid of models and showed that rotation-induced mixing is able to explain the large scatter of [Sr/Ba] observed in extremely metal-poor stars. Since there is still a number of stars at the lowest metallicities with only upper limits on Sr and/or Ba, increasing the sample sizes and the quality of the available high-resolution spectroscopy for stars at these metallicities is an essential step toward understanding nucleosynthesis at the earliest epochs and ultimately to characterize  environmental influence of the astrophysical sites of heavy elements production \cite{Roederer13}.

As already mentioned, the AGB phase represents the last stage of nuclear burning of small and intermediate mass stars. The AGB phase is short when compared to the MS stage, but is very important because it is a rich site of nucleosynthesis. These stars, once the supply of He for fusion in their core is exhausted, draw energy from fusion in the H and He shells around the degenerate C-O core. In this phase, the stars increase their brightness and their size, losing material from the outer layers due to strong stellar winds. One of the characteristics of the AGB phase is the intermittent thermal instability of the He-burning shells. These energy bursts manifest themselves as thermal pulses and hence this phase is known as the TP-AGB phase. These pulses typically happen every $10^{4}$-$10^{5}$ years (see, for more details, \cite{Herwig05}).
The s-process nucleosynthesis in AGB stars occurs in relatively low neutron density conditions ($\sim$ 10$^{7}$ neutrons/cm$^{3}$) during the late stages of stellar evolution when the star has a thin radiative layer (intershell region) and an expanded convective envelope (Fig. \ref{agbstar}). The main neutron enrichment sources are the $^{13}$C($\alpha$,n)$^{16}$O reaction, which releases neutrons radiatively during interpulse periods, and the $^{22}$Ne($\alpha$,n)$^{25}$Mg reaction, partially activated during the convective thermal pulses. 
The production of neutrons through the  $^{22}$Ne($\alpha$,n)$^{25}$Mg channel is really efficient only in high-mass AGB stars  ($M$ $\geq$ 4 $M_{\odot}$), due to the high temperature required for this reaction to occur. Such temperatures can also be reached during the TP-AGB phase of less massive stars, but in this case, the produced neutrons  only marginally affect the final distribution of the s-process abundances. This reaction takes place in a convective environment.
The $^{13}$C($\alpha$,n)$^{16}$O reaction requires proton and $\alpha$-capture reactions to occur at the same time in the He shell. One of the problems related to the modelling of this channel of neutron production is the low abundance of $^{13}$C. This element is produced during the phases following the development of the intermediate convective layer where some penetration of protons creates a reservoir of H in the He-rich layers. When the outer parts of the star re-contract and heat, H-burning is ignited again and the trapped protons are captured by the abundant $^{12}$C, inducing the chain $^{12}$C(p,$\gamma$)$^{13}$N($\beta$$^{+}$$\nu$)$^{13}$C. 
However, the $^{13}$C produced is not sufficient to explain the neutron production needed for the s-process nucleosynthesis. The reaction occurs in a radiative environment and leads to the formation of the so-called $^{13}$C $pocket$ (see \cite{Cristallo09} and references therein). The mass of the $^{13}$C $pocket$ is $\Delta$$M$ $\simeq$ 7 $\times$ $10^{-4}$ $M_{\odot}$ \cite{Busso99}, and the temperature required for this reaction is of the order of $T$ $\approx$ 9 $\times $ $10^{7}$ K. 
Recently, new calculations aiming at clarifying the $^{13}$C issue have been carried out \cite{Nordhaus08, Trippella14, Trippella16}. The models are based on the development of toroidal magnetic fields, induced by stellar dynamos, in the radiative He-rich layers below the convective envelope, and help to constrain the nucleosynthesis results obtained with the extension of the $^{13}$C $pocket$ with observations from the solar composition \cite{Trippella16}.

Recent advances in s-process nucleosynthesis are related to the determination of the neutron density in massive AGB stars \cite{D'Orazi13, Zamora14}. In particular, compared to solar abundances, the spectra of massive AGB stars in our Galaxy and the Magellanic Clouds, reveal a strong overabundance of Rubidium \cite{Hernandez06, Garcia09} and high [Rb/Zr] ratios \cite{Raai12}. Rb is an example of an element produced not only by the s-process but also by the r-process. The exact contribution of the two processes depends on the s-process model used to estimate the abundance, which is directly linked to the neutron enrichment process and, consequently, to the local neutron density. AGB star nucleosynthesis models \cite{Busso99} are far from matching the extreme Rb and [Rb/Zr] values, and the explanation of Rb overabundance could eventually lead to a better understanding of the $^{22}$Ne($\alpha$,n)$^{25}$Mg reaction. Within the framework of the s-procees, it is difficult to explain the lack of co-production of Zr, which is part of the same production peak of Rb and should be produced in similar quantities. 
Some solutions have been discussed in the literature to account for the Rb overabundance.
Karakas $et$ $al$ \cite{Karakas12} demonstrated that for solar-metallicity stars, [Rb/Fe] $\sim$ 1.4 could be reached if the final stage of mass loss was delayed, resulting in a larger number of thermal pulses and increased Rb production. However, the observed [Zr/Fe] ratios are roughly solar (within 0.5 dex \cite{Garcia07}), suggesting no production of this element in intermediate mass AGB stars. A different explanation has been proposed, such as the possibility that the gaseous Zr, having a condensation temperature (1741 K) \cite{Lodders03} greater than that of Rb (800 K), condenses in dust grains, producing an apparent lack of Zr, when measured using the molecular bands of ZrO \cite{Dinerstein06, Raai12}. Others possible solutions to the problem include the fact that the Magellanic Cloud observations are very uncertain, an incomplete understanding of the atmospheres of luminous AGB stars \cite{Garcia09} and a different AGB mass-loss rate \cite{Ventura05, D'Orazi13b}. Clearly, the future resolution of the rubidium problem promises to be an exciting challenge.

The nucleosynthetic model for the lighter s-process elements between Sr and Ba is not well understood yet. Travaglio $et$ $al$ \cite{Travaglio04} studied these elements, and by summing up all contributions from their model, the authors found that 8\%, 18\% and 18\% of Sr, Y, and Zr were missing. This missing fraction is assumed to come from massive primary origin stars at low [Fe/H] \cite{Travaglio04}. Because the process mainly affects the lighter peak elements this additional (unknown) nucleosynthetic contribution is called Lighter Element Primary Process (LEPP), or $weak$ r-process \cite{Ishimaru05}, and could explain some differences between these elements. Recently, the LEPP abundances have been further investigated by many authors \cite{Montes07, Izutani09, Bisterzo14} confirming the need for an additional process to account for the missing component of the light s-process isotopes. Most of the elements are produced through a mixture of s- and r-process \cite{Arlandini99}. This makes it harder to determine which of the processes are involved when creating the elements. 
The main component of the s-process is produced at metallicities starting at [Fe/H] $\sim$ -0.66 \cite{Cui07}, which corresponds to the time interval t $>$ 2.6 Gyr. Going to even lower metallicities or further back in time, gives insight into an undiluted view on other processes. At lower metallicities, from [Fe/H] $\sim$ -1.16 to [Fe/H] $\sim$ -0.66 \cite{Cui07}, the site for the strong component of the s-process was identified. At even lower metallicities, before the s-process sets in, the LEPP is believed to occur somewhere in stars. 
According to Cristallo $et$ $al$ \cite{Cristallo15} a variation of the standard paradigm of AGB nucleosynthesis would make it possible to reconcile  model predictions with solar system s-only abundances. However, the LEPP cannot be definitely ruled out, because of the uncertainties still affecting stellar and galactic chemical evolution models.
Several scenarios have been recently explored, both involving the primary r-process during the advanced phases of explosive nucleosynthesis (see \cite{Thielemann11} for a review) or a secondary s-process in massive stars (e.g., cs-component  \cite{Pignatari13}). Therefore, even if promising theoretical improvements related to the explosive phases of massive stars and ccSNe, as well as recent spectroscopic investigations \cite{Roederer12, Hansen12} have been made, a full understanding of the origin of the neutron capture elements from Sr up to Ba is still lacking.

New models and observations have suggested that in addition to the well-known slow- and rapid neutron capture processes, there may be an intermediate mode of neutron capture nucleosynthesis, the so-called i-process. This process is defined by a neutron flux larger than those found in the well-established s-process, yet smaller than the extreme conditions of the r-process. A possible signature of the i-process \cite{Cowan77} could be the simultaneous enhancement of Eu, usually considered an r-process element, and La, usually considered an s-process element, in some carbon enhanced metal-poor stars that have been classified as CEMP-r/s stars \cite{Dardelet15}. Post-AGB stars have been discussed earlier as possible nucleosynthesis sites for the i-process, yet there are still discrepancies and open questions to be addressed. In a new study by Jones $et$ $al$ \cite{Jones16}, super-AGB stars are identified as another possible astrophysical site for the i-process. In their new computational models of these very heavy AGB stars, mixing at convective boundaries are taken into account according to a parameterized model. These new stellar evolution models suggest that proton-rich material could be convectively mixed into the He-burning shell, leading to conditions suitable for the i-process. Interestingly, i-process conditions are more prominently found in models with lower metal content, indicating that the i-process could have been more important in the early universe. 1-D stellar evolution models can only identify possible sites for i-process nucleosynthesis \cite{Jones16} but  H-ingestion ashes are likely associated with substantial nuclear energy release, perhaps reaching  the level of the local binding energy of the He-burning shell. Such an enormous energy input is coupled with multi-scale turbulent mixing which cannot be realistically described with 1-D simulations \cite{Herwig14}.
3D stellar hydrodynamics simulations are mandatory to understand these nuclear astrophysical events fully and provide the appropriate context for further investigations.

The possibility of neutrons catalyzing the formation of heavier nuclei was recently proposed. The formation of the Rydberg nuclear molecule $^{16}$O ($^{10}$Be + n + n + $^{10}$Be) for example, might exist in rich neutron environments within AGB stars \cite{Frederico17}. In this mechanism of formation, neutrons mediate the Efimov long-range interaction of the Be nuclei, and could eventually be used to form other nuclear molecules with heavier nuclei, facilitating a nuclear reaction and eventually nucleosynthesis. Calculations show that one cannot confirm this possibility. But it is also difficult to rule out the existence of such molecules based on what is known about nuclear interactions.

\subsection{Supernovae}
\label{sec:Supernovae}

Supernovae can be classified into two basic kinds: Type Ia (SNIa) which are thought to be the explosion of a WD in a binary system that accretes sufficient mass from its companion, and all the rest (Type II, Ib, Ic), which are generated within several possible scenarios (for a review of all scenarios see, \cite{Lapuente14, Maoz14, Woosley05}). Observationally they can be  classified according to the absence (Type I) or presence (Type II) of H lines in their spectra. Type II (SNII), Ib and Ic, are produced from massive stars $\approx$ 10 $M_{\odot}$ and are observed in the spiral and irregular galaxies. SNIa happens in all types of galaxies with no preference for star-forming regions, consistent with their origin in old or intermediate age stellar populations.

\subsubsection{Nucleosynthesis in Type Ia supernovae}

SNIa are important nucleosynthesis sites for iron group elements and possibly the p-process. 
Within the framework of SNIa the general scenario is that a C-O WD  accretes mass from a companion star in a binary system until it ignites near  the Chandrasekhar mass \cite{Hoyle60}. The companion star of the C-O WD is usually a He-burning star or a He-rich WD \cite{Wang09, Guillochon10, Ruiter14, Wang15, Geier15}. It has been proposed that the detonation of the He-rich shell is triggered via thermal instability if the companion of the C-O WD is a He star (e.g., \cite{Nomoto82}), whereas the detonation of the He-rich envelope is ignited dynamically if the companion is a He-rich WD (e.g., \cite{Guillochon10}). For more discussion on the progenitors of SNIa see, \cite{Chen12, Liu12, Wang13, Zhou16}.
WD instabilities are relevant for SNIa since they are related not only to strong magnetic fields in the star interior \cite{Coelho14} but also to neutronization due to electron capture reactions. Because of this reaction, atomic nuclei become more neutron-rich and the energy density of the matter is reduced, at a given pressure, leading to a softer equation of state. Other nuclear reactions that make very massive WDs unstable are the pycnonuclear fusion reactions in the cores of these compact stars \cite{Boshkayev13, Otoniel16}. These reactions among heavy atomic nuclei, schematically expressed as $^{A_{i}}_{Z_{i}}Y_{i} + ^{A_{j}}_{Z_{j}}Y_{j} \rightarrow ^{A_{i}+A_{j}}_{Z_{i}+Z{j}}Y_{k}$, are possible due to the high density matter of  WDs; an important reaction is carbon on carbon, $^{12}\textrm{C}+^{12}\textrm{C}$. Pycnonuclear reactions have been found to occur over a significant range of stellar densities (see, for instance, \cite{Gasques05}), including the density range found in the interiors of WDs \cite{Chamel13, Chamel15}. Recently, WD calculations also showed that central energy densities are limited by nuclear fusion reactions and inverse $\beta$-decay \cite{Chamel13, Boshkayev13}. The nuclear fusion rates at which very low energy pycnonuclear reactions proceed, however, are highly uncertain because of  poorly constrained parameters \cite{Yakovlev06}. Finally, it should be mentioned that very recently it has been suggested that pycnonuclear reactions could be able to drive powerful detonations in single C-O WDs \cite{Chiosi15}.


\subsubsection{Core-collapse supernovae: observations}

Explosive nucleosynthesis is associated with the passage of the ccSN shock wave through the layers above the PNS (see \cite{Woosley02} for a review). The shock heats the matter it traverses, inducing an explosive nuclear burning characterized by short times, which leads to large deviations from equilibrium and hydrostatic nuclear burning patterns. This explosive nucleosynthesis can alter the elemental abundance distributions in the inner (Si, O) shells.
The properties of the process are tied to those of the explosion. 
The details of the nucleosynthesis, which produces radioactive nuclei such as $^{26}$Al, $^{28}$Si, $^{44}$Ti, $^{56}$Ni and $^{56}$Co during the explosions, are not yet fully understood. In order to shed light on the mechanism which drives the explosive nucleosynthesis, one obtains hints from the observations of the energy and material which are injected into the interstellar medium from ccNS explosions. Some of this material, which is the result of nucleosynthesis processes occurring during the explosion, is made of radioactive isotopes, and therefore allow us to infer the ccSN nucleosynthesis conditions which are needed in order to produce them. For instance, the observations of gamma-rays from $^{44}$Ti and $^{56}$Ni in ccSN events represents a valuable tool to penetrate deeply into the interior of these explosions, which are otherwise only accessible through neutrinos \cite{Wongwathanarat16}. In this section, we outline the results of the comparison of explosive nucleosynthesis models with observations from ccSNe.
Since the launch of the INTEGRAL observatory, it has been possible to accurately determine the gamma-ray flux associated with heavy elements produced by astrophysical sources. 
The main site of production of the radioisotope $^{44}$Ti is thought to be the innermost ejected layers of ccSN explosions, 
and the study of its abundance has been the focus of several works \cite{Magkotsios10, Nagataki98, Woosley91}.
The $^{44}$Ti yield of ccSNe is notoriously difficult to calculate because it depends on the explosion energy and on the symmetry of the explosion \cite{Weidenspointner06}. Theoretical calculations indicate that both increased explosion energy and increased asymmetry result in an increased $^{44}$Ti yield. Observationally, the presence of the radioisotope $^{44}$Ti is revealed to the gamma-ray astronomer through the emission of three gamma-ray lines. The decay $^{44}$Ti $\rightarrow$ $^{44}$Sc gives rise to gamma rays at 67.9 keV and 78.4 keV. The subsequent decay $^{44}$Sc $\rightarrow$ $^{44}$Ca gives rise to a line at 1157.0 keV. The amount and the velocity of $^{44}$Ti is a powerful probe of the explosion mechanism and dynamics of ccSNe, and in addition, the $^{44}$Ti gamma-ray line emission is an ideal indicator of young supernovae remnants. 
Up to now, $^{44}$Ti has not yet been directly detected in SN 1987A. From modelling of the ultraviolet optical infrared (UVOIR) light curves  based on the radioactive decays, different values of the amount of produced $^{44}$Ti have been predicted, which do not always agree with each other within the respective uncertainties. For example, from the analysis of X-ray data taken from INTEGRAL, Grebenev $et$ $al$ \cite{Grebenev12} suggested a value of (3.1 $\pm$ 0.8) $\times$ $10^{-4}$ $M_{\odot}$, while UVOIR bolometric light curve analysis of Seitenzahl $et$ $al$ \cite{Seitenzahl14} indicate a value (0.55 $\pm$ 0.17) $\times$ $10^{-4}$ $M_{\odot}$. 
Furthermore, there is a lack of consistency between the theoretical predictions and observations. Spherically symmetric (1D) models of SN 1987A produce, in general, a few $10^{-5}$ $M_{\odot}$ of $^{44}$Ti \cite{Seitenzahl14}. Perego $et$ $al$ \cite{Perego15} used the method PUSH to produce a 1D supernova explosion, that better fits the produced amounts of $^{56}$Ni in SN 1987A, predicting an amount of 3.99 $\times$ $10^{-4}$ $M_{\odot}$ for $^{44}$Ti.
Magkotsios $et$ $al$ \cite{Magkotsios10} investigated the $^{44}$Ti abundance produced from ccSNe by studying the impact on the $^{44}$Ti abundance evolution of variations of the nuclear reactions, including ($\alpha$,$\gamma$), ($\alpha$,p), (p,$\alpha$), and ($\alpha$,n) reactions in light and intermediate mass targets. They found that the variation in the $^{17}$F($\alpha$,p)$^{20}$Ne reaction rate causes the primary impact on the $^{44}$Ti abundance. The $^{17}$F($\alpha$,p)$^{20}$Ne reaction rate, however, has never been measured. Because the reaction rate may be dominated by the properties of energy levels of $^{21}$Na above the $\alpha$-threshold at 6.561 MeV, searching for energy levels of $^{21}$Na and studying their properties may have impact on our understanding of the abundance evolution of $^{44}$Ti. In this context, the reaction $^{24}$Mg(p,$\alpha$)$^{21}$Na plays a central role and knowledge of its rate is of key importance. The $^{24}$Mg(p,$\alpha$)$^{21}$Na reaction was recently measured by Cha $et$ $al$ \cite{Cha15} in order to make a spectroscopic study of the energy levels in  $^{21}$Na for the $^{17}$F($\alpha$,p)$^{20}$Ne reaction rate at stellar temperatures. Further comparisons between observations and models are clearly required in the future, and more sensitive hard X-/soft-gamma ray astronomical instrumentations are required \cite{Frontera13}.

The short-lived radioisotope $^{56}$Ni is also synthesized in the deep interiors of ccSN explosions. CcSN light is understood as being powered mainly by $^{56}$Ni radioactive decay, as demonstrated by the characteristic light curve and spectral-evolution data \cite{Colgate69}. 
These radioactive isotopes carry information about the environment of the explosion formation, unaffected by the violent expansion of the ccSN \cite{Diehl13}. 
One of the key issues from observations is the broad range of inferred amounts of $^{56}$Ni. The proximity of SN 1987A allowed the very first detection of gamma-ray lines from the radioactive process $^{56}$Ni $\rightarrow$ $^{56}$Co $\rightarrow$ $^{56}$Fe \cite{Matz88}. By estimates of the gamma absorption towards the supernova remnant Cas A and the abundance of Fe  from X-ray observations, Eriksen $et$ $al$ \cite{Eriksen09} predict the abundance of $^{56}$Ni to be in the range (0.58-0.16) $M_{\odot}$. The standard value is $^{56}$Ni $\sim$ 0.07 $M_{\odot}$ \cite{Seitenzahl14}. Different theoretical predictions have been carried out for the amount of $^{56}$Ni. Concerning Cas A, Magkotsios $et$ $al$ \cite{Magkotsios10} post-processed the trajectories of a 1D ccSN model from Young $et$ $al$ \cite{Young08}, whose progenitor was designed to match Cas A, and obtained a value of 2.46 $\times$ $10^{-1}$ $M_{\odot}$ for $^{56}$Ni. Using a two-dimensional rotating 15 $M_{\odot}$ model of Fryer and Heger \cite{Fryer00}, they obtain a higher value of 3.89 $\times$ $10^{-1}$ $M_{\odot}$ for $^{56}$Ni. However, it must be emphasized, that all the above models do not follow the ccSN shock wave long enough, and therefore hydrodynamic trajectories have to be extrapolated in order to be able to perform nucleosynthesis calculations.

Another important element which is synthesized during the final burning stage is the gamma-ray emitter $^{26}$Al, which has been detected in the interstellar medium of our Galaxy \cite{Smith04, Wang07}. $^{26}$Al is produced mainly in massive star winds and during ccSN explosions. $^{26}$Al production for different candidate sources has been estimated by various
groups \cite{Tur10, Limongi06, Diehl06}.  Chieffi and Limongi \cite{Chieffi13} include stellar rotation and its effect upon the computed yields compared to non-rotational models. The galactic mass yields of $^{26}$Al is $\sim$ 1.7-2.0 $\pm$ 0.2 $M_{\odot}$ \cite{Martin09}.
Voss $et$ $al$ \cite{Voss10} studied the variations between different models of massive stars, in particular, the effects of rotation and the strength of wind mass-loss on the radioactive tracers and the energetics of star-forming regions. The individual nearby star-forming regions Sco-Cen \cite{Diehl10}, Orion \cite{Voss10}, and Cygnus \cite{Martin10} have been studied in detail and good agreement has been found between theory and observations.
Theoretical ccSN models, however, suffer from considerable uncertainties in $^{26}$Al production because of a lack of experimental knowledge of the reactions that create and destroy $^{26}$Al under ccSN conditions \cite{Woosley95, Limongi06}. For instance, the uncertainties in the nuclear reaction rates responsible for the formation of  $^{26}$Al ejected in the supernova explosions lead to uncertainties of a factor $\sim$ 3 in its abundance \cite{Iliadis11}. 
Classical novae \cite{Starrfield16} are one potential source of $^{26}$Al and it has been shown that up to 0.4 $M_{\odot}$  of the galactic abundance could have been produced in these sites \cite{Jose97}.
In particular, the $^{26}$Al(p,$\gamma$)$^{27}$Si \cite{Ruiz06} reaction strongly affects the abundance of $^{26}$Al in nova ejecta. The short-lived isomer, $^{26m}$Al plays a special role  in novae, since $^{26m}$Al and $^{26}$Al are in quasi-equilibrium in these conditions, and thus knowledge of both, the destruction of the ground state and isomer, is needed to determine the effective $^{26}$Al half-life and ejected abundance. Indirect studies could help to determine the $^{26m}$Al(p,$\gamma$)$^{27}$Si reaction rate. Moreover, the $^{23}$Mg(p,$\gamma$)$^{24}$Al reaction also contributes to $^{26}$Al  nucleosynthesis in novae. The $^{23}$Mg(p,$\gamma$)$^{24}$Al reaction was measured directly for the first time at the DRAGON facility to a precision sufficient for novae yield purposes \cite{Erikson10}. Measurements led to a reduction in the uncertainties of ejected $^{26}$Al in the types of nova model seen in, e.g., \cite{Iliadis02}. Nevertheless, at temperatures lower than those reached in O-Ne classical novae, the rate is still dominated by direct capture and uncertainties will be related to this component.

\subsubsection{r-process}

Neutrinos play a crucial role in our understanding of SNII (see, for example, \cite{Roberts16}). According to the currently most widely accepted theory for the explosion of a massive star, the explosion energy is provided by the neutrinos that are abundantly emitted from the nascent PNS and interact with the material of the progenitor star (Fig. \ref{fig:driven}). This energy deposition is not only supposed to power the propagation of the supernova shock into the stellar mantle and envelope regions, as well as to cause the violent disruption of the star, but also drives a mass outflow from the surface of the PNS. This continues for more than 10 seconds and might be a suitable site for r-process nucleosynthesis. The baryonic outflow that expands with supersonic velocities  is known as the neutrino-driven wind \cite{Arcones13}. The PNS cools by emitting neutrinos, i.e., $\nu_{e}$, $\bar{\nu}_{e}$. As these neutrinos pass through the hot material predominantly consisting of free nucleons immediately outside the PNS, a fraction of the $\nu_{e}$ and $\bar{\nu}_{e}$ can be absorbed through $\nu_{e}$ + $n$ $\rightarrow$ $p$ + $e^{-}$ and $\bar{\nu}_{e}$ + $p$ $\rightarrow$ $n$ + $e^{+}$. On average, a nucleon obtains $\sim$ 20 MeV from each interaction with $\nu_{e}$ or $\bar{\nu}_{e}$. In order to escape from the PNS gravitational potential of $G$$M_{NS}$$m_{u}$/$R_{NS}$ $\sim$ 200 MeV, a nucleon in the wind must interact with $\nu_{e}$ and $\bar{\nu}_{e}$  $\sim$ 10 times. Eventually, the neutrino-driven wind collides with the slow, early ccSN ejecta resulting in a wind termination shock or reverse shock \cite{Fischer10}. The above reactions also interconvert neutrons and protons, thereby determining the electron fraction $Y_{e}$ in the wind \cite{Qian93}. The neutrino-driven wind has attracted vast attention over the last 20 years as it was suggested to be a candidate for the astrophysics site where half of the heavy elements are produced via the r-process \cite{Arcones13}. The general conditions required for the r-process have been investigated both via analytical \cite{Qian96} and via steady-state \cite{Thompson01} models of neutrino-driven winds.

\begin{figure}
\resizebox{0.48\textwidth}{!}{%
  \includegraphics{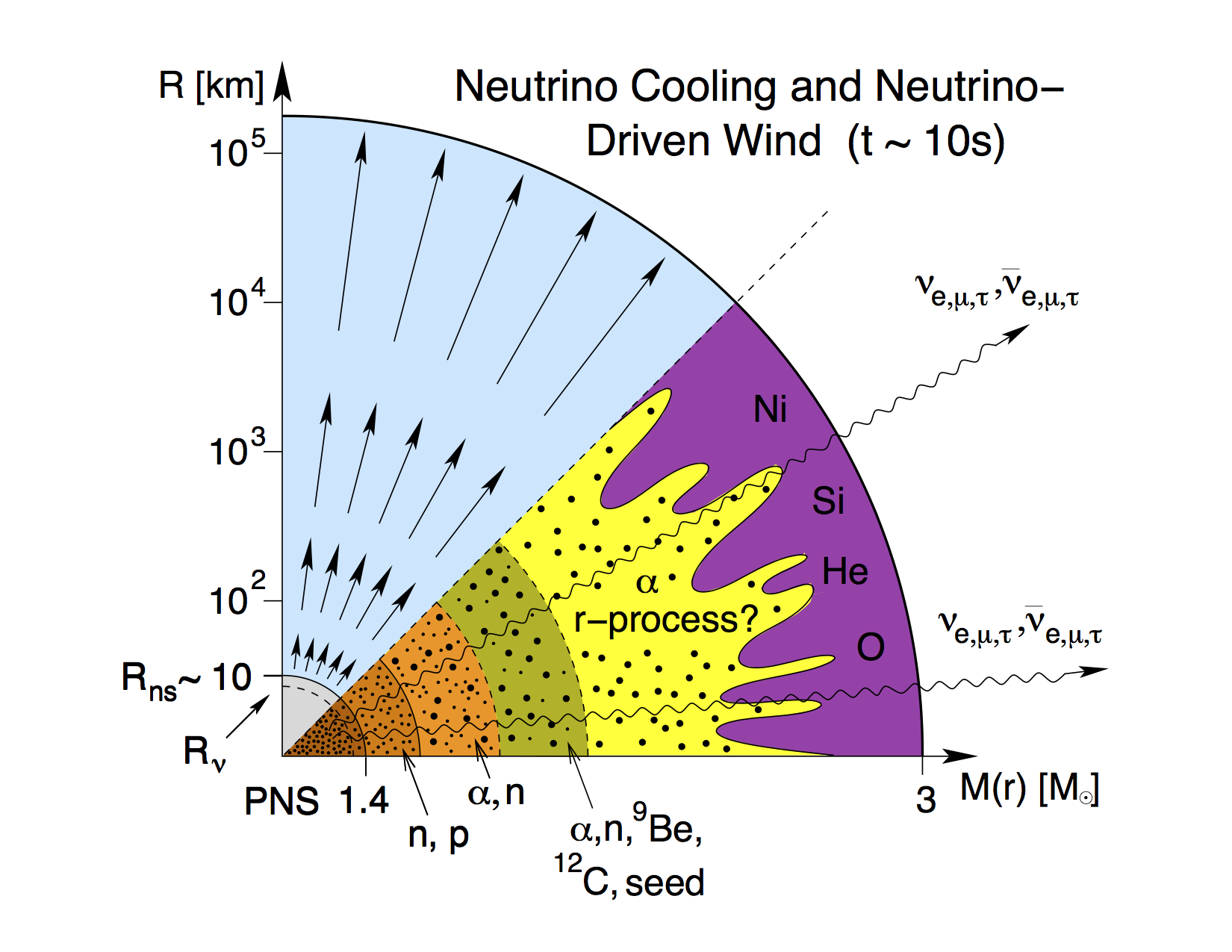}
}
\caption{Schematic representation of the neutrino-driven wind from the surface of a recently born PNS. The horizontal axis gives mass information whereas the vertical axis shows corresponding radii, with $R$\textsubscript{ns} and $R_{\nu}$  being the NS and neutrinosphere, respectively. The supersonic neutrino-driven wind, which forms after the onset of the explosion, provides favorable conditions for the r-process -- a high neutron abundance, short dynamical time scales, and high entropies. In this high-entropy environment, it is possible that most nucleons are in the form of free neutrons or bound into alpha particles. Thus, there can be many neutrons per seed nucleus even though the material is not particularly neutron-rich. The predicted amount of r-process material ejected per event from this environment agrees well with that required by simple galactic evolution arguments. For more details, the reader is referred to \cite{Janka07}.}
\label{fig:driven}       
\end{figure}

In order to account for the solar r-process abundances associated with the peaks at $A$ $\sim$ 130 and 195, each supernova must eject $\sim$ $10^{-6}$--$10^{-5}$ $M_{\odot}$ of r-process material. Although the current neutrino-driven wind models have difficulty in providing conditions for heavy r-process  \cite{Hudepohl10}, the wind naturally ejects $\sim$ $10^{-6}$-$10^{-5}$ $M_{\odot}$ of material over a period of $\sim$ 1 s \cite{Qian03}. This is because the small heating rate due to the weakness of neutrino interaction permits material to escape from the deep gravitational potential of the PNS star at a typical rate of $\sim$ $10^{-6}$--$10^{-5}$ $M_{\odot}$ s$^{-1}$ \cite{Thompson01}. Indeed, the ability to eject a tiny but interesting amount of material was recognized as an attractive feature of the neutrino-driven wind model of the r-process (e.g., \cite{Meyer92}).
However, current models fail to provide the conditions for an r-process to occur in the wind. For instance, the production of heavy r-process elements ($A$ $>$ 130), requires a high neutron-to-seed ratio. This can be achieved by the following conditions: high entropy, fast expansions or low electron fraction \cite{Otsuki00, Thompson01}. As Arcones $et$ $al$ \cite{Arcones11} remarked, these conditions are not yet realized in hydrodynamical simulations that follow the outflow evolution during the first seconds of the wind phase after the explosion \cite{Arcones07}. 
Contrarily, the $weak$ r-process, which accounts for the lighter neutron-capture elements ($A$ $\sim$ 80 peak), is strongly believed to take place in neutrino-driven winds that could occur in ccSNe or collapsar accretion disks \cite{Wanajo06}. The astrophysical conditions required to produce the peak region 
 through weak r-process can be found in the recent study by Surman $et$ $al$ \cite{Surman14}. Once the wind has cooled down after a few seconds, charged particle reactions are key to the production of the heavy elements. For a typical wind evolution, the ($\alpha$,n) reaction is faster than all other charged particle reactions, thus driving nucleosynthesis evolution in neutron-rich winds. None of the most relevant ($\alpha$,n) reactions had been measured in the energy range relevant for weak r-process astrophysical conditions. So far modelers have to rely on theoretical predictions of those rates. Furthermore, the theoretical uncertainties of the calculated reaction rates can be as high as 2 orders of magnitude and abundance network calculations are highly sensitive within the expected theoretical uncertainties of these rates \cite{Pereira16}. A recent systematic study searching for the critical reaction rates that influence most of the final abundances in weak r-process scenarios has permitted the identification of the most important reaction rates, which can then be pinned down experimentally by measurements in radioactive beam facilities \cite{Bliss17}. Most of the reaction rates responsible for the production of elements ($A$ $\sim$ 80) in neutrino-driven winds are either viable with current beam intensities in existing nuclear physics facilities or will be in the near future.  Nuclei that participate in the r-process typically have half-lives that are too short to allow them to be made into a target. As neutron targets are not available, neutron capture experiments performed on these nuclei represent a big challenge. Improvements in theoretical reaction rates are needed, along with advances in experimental facilitates, to reduce these fundamental nuclear physics uncertainties. Another possible scenario could be the r-process nucleosynthesis in the neutrino-driven outflows from the thick accretion disk (or "torus") around a BH, as recently investigated by Wanajo $et$ $al$ \cite{Wanajo12}. A BH accretion torus is expected as a remnants of a binary NS or NS-BH merger. The computed mass-integrated nucleosynthetic abundances are in good agreement with the solar system r-process abundance distribution, suggesting that BH torus winds from compact binary mergers have the potential to be a major, and in some cases dominant, production site of r-process elements \cite{Wu16}.

There is direct evidence that ccSNe also produce Magnetohydrodynamic (MHD) jets with a power comparable to the explosion itself \cite{Meier76, Wheeler02, Mosta14}. Expected speeds are $\sim$ 0.25-0.5c (the escape speed from the new PNS). While NSs are expected to be left after ccSN explosions, it has been suggested that a star of more than 25 $M_{\odot}$ may collapse to a BH \cite{Heger03}; an accretion disk is formed around the BH if the star has enough angular momentum before the collapse. This system could produce a relativistic jet of gamma-rays  (GRBs, see Section \ref{sec:gamma}) due to MHD effects, a system  called a collapsar model \cite{Woosley93}.
Magnetically driven jets of collapsar models have been extensively investigated as a site of the r-process \cite{Fujimoto06, Nishimura06}. Strong magneto-rotational driven jets of the collapsar model can produce heavy r-process nuclei with a very simple treatment of BH formation \cite{Ono12}. Estimates of the composition of jets ejected by a collapsar have shown that the synthesis of heavy elements can occur also in the ejection phase during the core-collapse of the star \cite{Fujimoto07}. It was found that elements like U and Th are synthesized through the r-process when the source has a large magnetic field (10$^{12}$ G). In addition, many p-nuclei are produced in the jets. The material far from the axis does not fall straight in but forms an accretion disk first, if the angular momentum of the star is high enough. For high accretion rates, the accretion disk is so dense and hot that nuclear burning is expected to proceed efficiently, and the innermost region of the disk becomes neutron-rich through electron captures on nuclei. This region is an efficient r-process site, and about 0.01 $M_{\odot}$ of massive neutron-rich nuclei can be ejected from the collapsar,  U and Th being the most abundantly synthesized elements \cite{Nakamrua12}. Recent nucleosynthesis calculations in a three-dimensional MHD supernova model have suggested that such supernova could be the source of the r-process elements in the early galaxy \cite{Winteler12}. However, in those calculations, the produced nuclei are limited to primary synthesized ones inside the jets and comparisons with the solar system abundances have been focused on elements heavier than the iron-group nuclei. Ono $et$ $al$ \cite{Ono09, Ono12} performed explosive nucleosynthesis calculations inside the jetlike explosions for the collapsar of a massive He core star of 32 $M_{\odot}$. These calculations include hydrostatic nucleosynthesis using a nuclear reaction network, which has 1714 nuclei (up to $^{241}$U). The jet model cannot  produce both the elements around the third peak of the solar r-elements and intermediate p-elements when compared with the previous study \cite{Fujimoto07, Fujimoto08} of r-process nucleosynthesis calculations in a collapsar model of 40 $M_{\odot}$. This may be attributed to the differences in the progenitor and the specified initial angular momentum and magnetic field distributions.
A study by Banerjee $et$ $al$ \cite{Banerjee13} has shown that the synthesis of rare elements, such as $^{31}$P, $^{39}$K, $^{43}$Sc, and $^{35}$Cl and other uncommon isotopes, is also possible. These elements, which are produced in the simulations in the outer regions of low \.{M} accretion disks (i.e., 0.001-0.01 $M_{\odot}$ s$^{-1}$), have been discovered in the emission lines of some long GRBs afterglows. However, they have yet to be confirmed by further observations. Still other models have been proposed. The list includes calculations based upon a model for a MHD+neutrino-heated collapsar jet \cite{Nakamura15}, prompt-magnetic-jet and delayed-magnetic-jet explosion models \cite{Nishimura15} and rapidly rotating strongly magnetized core-collapse models \cite{Mosta14, Mosta17, Halevi18}.
For additional information on topics related to the r-process in ccSNe, see \cite{Farouqi10}.

\subsubsection{p-process}
\label{sec:p-nuc}

In this section, we will outline the latest progress associated with the production of p-nuclei during supernovae explosions. A number of proton-rich isotopes cannot be synthesized through sequences of only neutron captures and $\beta$-decays, therefore  the postulation of a third process is required (see, for example, \cite{Meyer94} and references therein). There are several possibilities to get to the proton-rich side. As discussed above, p-nuclei are synthesized by successively adding protons to a nuclide or by removing neutrons from pre-existing s- or r-nuclides through sequences of photodisintegrations. Under conditions encountered in astrophysical environments, it is difficult to obtain p-nuclei through proton captures because the Coulomb barrier of a nucleus increases with increasing proton number. Furthermore, at high temperature ($\gamma$,p) reactions become faster than proton captures and prevent the build-up of proton-rich nuclides. Photodisintegration is an alternative way to make up p-nuclei, either by destroying  their neutron-richer neighbour isotopes through sequences of ($\gamma$,n) reactions or by flows from heavier and unstable nuclides via ($\gamma$,p) or ($\gamma$,$\alpha$) reactions and subsequent $\beta$-decays. It is clear that the term p-process is used for any process synthesizing p-nuclei, even when no proton captures are involved. Indeed, so far it seems to be impossible to reproduce the solar abundances of the p-isotopes by one single process. 
In our current understanding, there is evidence that more than one process in more than one astrophysical scenario is relevant for the production of p-nuclei \cite{Wanajo06, Frohlich06, Pruet06, Rauscher10, Kusakabe11}. Arnould \cite{Arnould76} proposed the p-process in presupernova phases, and Woosley and Howard \cite{Woosley78} proposed the $\gamma$-process in supernovae. This so-called $\gamma$-process requires high stellar plasma temperatures and occurs mainly in explosive O/Ne-burning during a ccSN (see, for example, \cite{Rapp06, Rauscher10, Pignatari16}). The $\gamma$-process during a ccSN explosion is the most well-established astrophysical scenario for the nucleosynthesis of  p-nuclei \cite{Woosley78}. Already in earlier works \cite{Rayet90, Rayet95} the O/Ne-rich layers of massive stars were considered to host the $\gamma$-process. The $\gamma$-process is activated with typical timescales of less than a second when the shock front passes through the O/Ne-burning zone.  Historically, 35 p-nuclides have been identified, with $^{74}$Se being the lightest and $^{196}$Hg the heaviest. The isotopic abundances of p-nuclei are 1-2 orders of magnitude lower than for the respective r- and s-nuclei in the same mass region. 
The nuclear reactions occurring in the $\gamma$-process are mainly induced by photons in the MeV-energy range,  the reaction rate being determined by the Planck distribution. Temperatures of the order of several $10^{9}$ K are required to provide sufficient energy. Such temperatures are realized within ccSN explosions. Explosive events also provide the correct timescale of several seconds -- if the photon intensity were to last for a longer period, the seed distribution would completely convert to light isotopes without leaving p-nuclei behind. In the early work of Woosley and Howard \cite{Woosley78},  different conditions were found to be required to produce the complete range of p-nuclei from  $^{74}$Se up to  $^{196}$Hg. Therefore, different density and temperature profiles were dedicated to different layers of material of the ccSNe. A typical range of peak temperatures is 2 to 3 $\times$ $10^{9}$  K while maximum densities vary between 2 $\times$ $10^{5}$ g cm$^{-3}$ and 6 $\times$ $10^{5}$ g cm$^{-3}$. A combination of a density profile and a temperature profile is often referred to as $trajectory$. These trajectories vary significantly for different astrophysical sites fulfilling the general conditions.

\begin{figure}
\begin{center}
\resizebox{0.48\textwidth}{!}{%
  \includegraphics{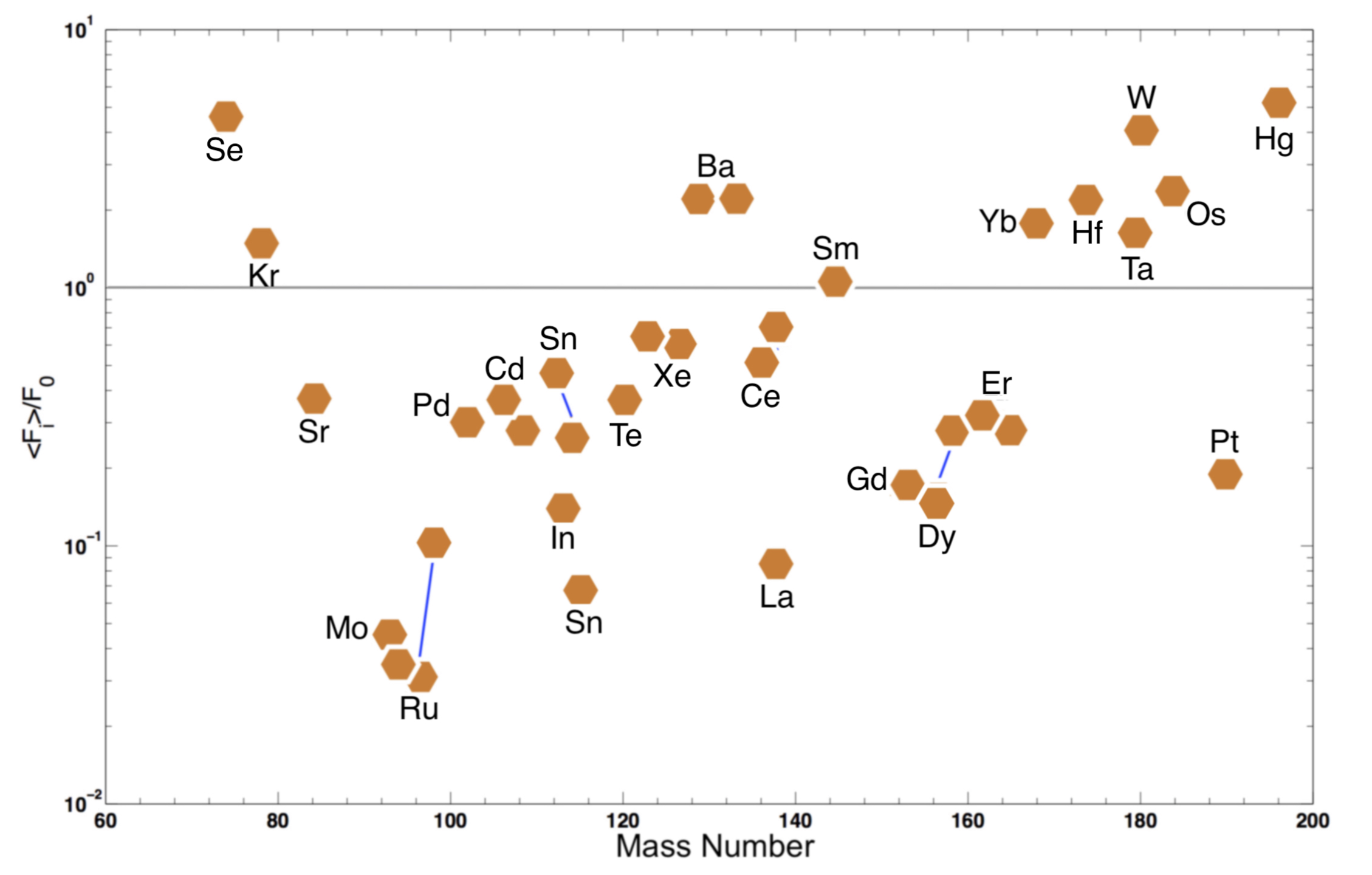}  
}
\resizebox{0.48\textwidth}{!}{%
\includegraphics{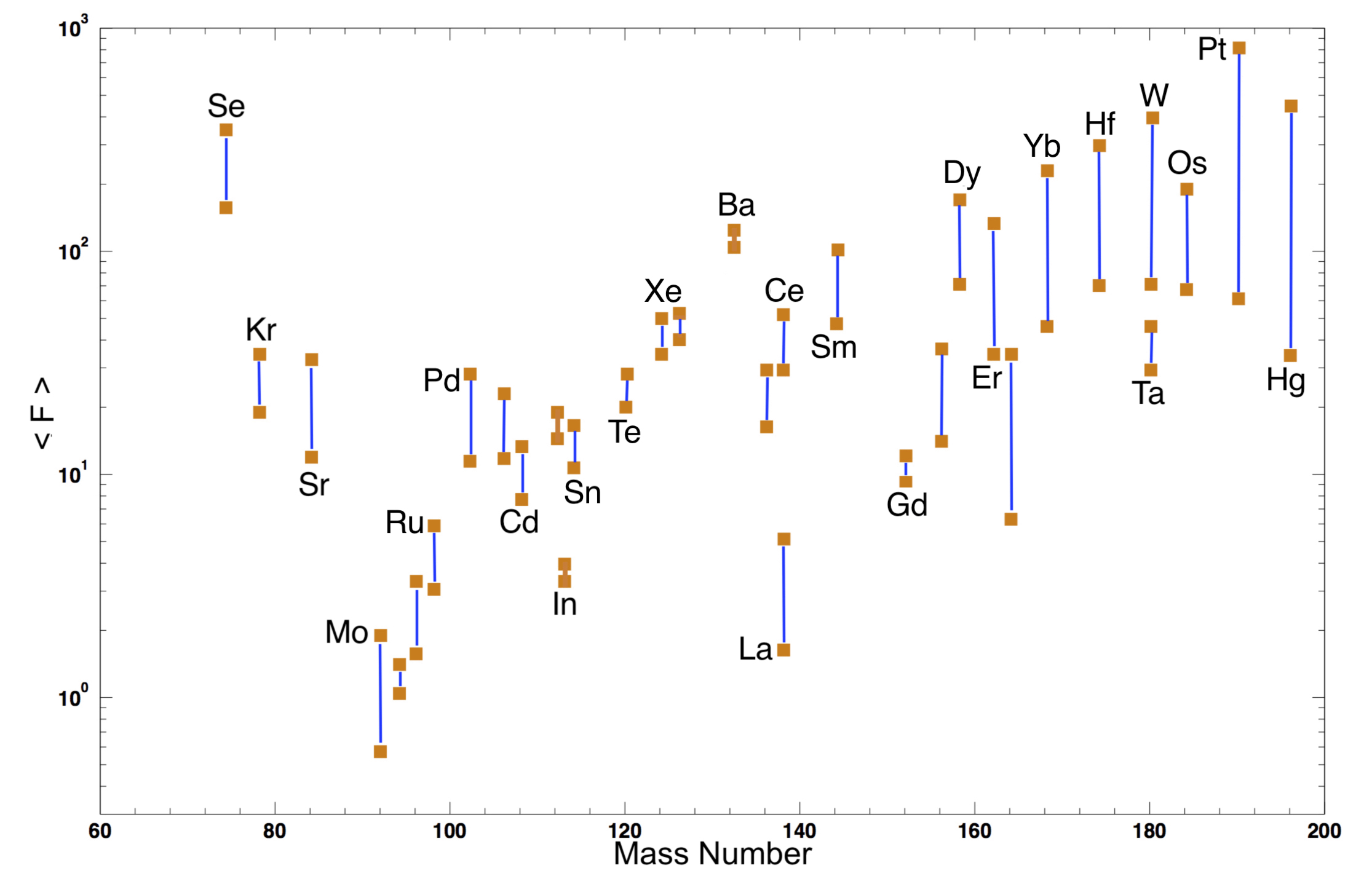}
}
\caption{Overproduction factors $<F>$ of p-nuclei in SNII of 25 $M_{\odot}$ stars. $Top$: The light p-nuclei $^{92,94}$Mo and $^{96}$Ru are the most strongly underproduced. Data taken from \cite{Rapp06}. $Bottom$: Displayed for each nuclide are the maximum and minimum abundances predicted from p-process calculations. The abundances of the lighter species are mainly affected by uncertainties in the predicted nuclear level densities and the nucleon-nucleus potential. Adapted from \cite{Arnould03}. For details, see text.}
\label{fig:over}  
\end{center}     
\end{figure}

It has been shown that the $\gamma$-process scenario suffers from a strong underproduction of the most abundant p-isotopes, $^{92,94}$Mo (see, for example, \cite{Fisker09}) and $^{96,98}$Ru. 
In contrast to the r- and s-process, the abundances produced in the $\gamma$-process vary significantly with the composition of the seed distribution. Detailed studies performed by Costa $et$ $al$ \cite{Costa00} showed that an enrichment of weak s-process material allows for a sufficient production of the Mo and Ru p-nuclei. At the same time, the overproduction factors of the lighter p-nuclei are further increased. Therefore, a variation of the seed distribution alone cannot solve the overabundance of the Mo-Ru isotopes. CcSN models cannot reproduce the relatively large abundances of $^{92,94}$Mo and $^{96,98}$Ru, even taking into account nuclear uncertainties  \cite{Rapp06, Rauscher06}, except for a possible increase of the $^{12}$C+$^{12}$C fusion reaction rate \cite{Pignatari13}. Based on the observations from metal-poor stars of the galactic halo, these elements can be considered as highly mixed elements, where contributions from s-process of stellar nucleosynthesis and main and weak r-processes are all mixed with smaller contributions from the main p-process. Alternatively, other processes in massive stars different from the classical p-process have been proposed to contribute to the missing Mo-Ru p-abundances, e.g., the $\nu$p-process in proton-rich neutrino wind conditions \cite{Rauscher13}. Mo and Ru are promising elements for studying the extent of planetary ­scale nucleosyn­thetic isotopic heterogeneity in the inner solar system. Both the elements have seven isotopes of roughly equal abundance that are produced by distinct nucleosynthetic processes. Furthermore, they occur in mea­surable quantities in almost all meteorite groups, permitting a comprehensive assessment of the extent of any isotopic heterogeneity in the inner solar system. Identifying isotope anomalies at the bulk meteorite scale provides important information regarding the extent and efficiency of mixing processes since the isotope variations are most readily accounted for by variable abundances of p-­, s-­ and r-­process in these samples.
Isotopic heterogeneity in iron meteorites and bulk chondrites has been observed for a number of elements, including Mo \cite{Dauphas02} and Ru \cite{Chen10}. These results contrast with evidence for isotopic homogeneity \cite{Becker03, Becker03b}. Mo isotopic anomalies in the bulk meteorites correlate with those in Ru exactly as predicted from nucleosynthetic theory, providing strong evidence that the correlated Ru and Mo anomalies are caused by a heterogeneous distribution of one or more s-process carriers \cite{Dauphas04, Chen10, Burkhardt11}. However, the extent of the isotope anomalies in meteorites is poorly constrained because previous studies ob­tained different results regarding the presence of Mo isotopic anomalies in meteorites \cite{Becker03, Dauphas02, Yin02}. The origin and extent of nucleosynthetic Mo-Ru isotope variations in meteorites and their components need to be further investigated and more detailed neutron capture process yields are required to determine their contribution to the abundance of the elements.
Typical theoretical overproduction factors are shown in Figure \ref{fig:over} for all p-nuclei. If the lightest p-nuclei $^{74}$Se and $^{80}$Kr are ignored, on average, a monotonic increase is observed with increasing mass number. This trend cannot be corrected by nuclear physics uncertainties as shown in \cite{Arnould03} but is based on the model, e.g., the heaviest p-nuclei only survive in the outermost layers with the lowest peak temperatures, an effect which might be overestimated in the current models. Usually, the seed composition is a mixture of r- and s-process nucleosynthesis as found in the solar abundance distribution. There are many excellent papers on the Mo-Ru problem, and the interested reader will find more information in \cite{Peterson13, Hansen14b, Poole17}.  

Another process in ccSNe that can produce the light p-process nuclei up to Pd-Ag, including $^{92}$Nb, is the combination of $\alpha$, proton, neutron captures, and their reverse reactions in $\alpha$-rich freezeout conditions \cite{Woosley92}. Neutrino winds from the forming NS are also a possible site for the production of the light p-process nuclei \cite{Farouqi09, Arcones11b}, although one of its possible components, the $\nu$p-process \cite{Frohlich06}, cannot produce  $^{92}$Nb because it is shielded by $^{92}$Mo \cite{Rauscher13}. The same occurs in the case of the rp-process in X-ray bursts \cite{Dauphas03} (see Section \ref{sec:rp-nuc}). Moreover, the total amount of p-nuclei produced in one event and the expected rate of SNII explosions do not match the absolute observed abundances. Therefore, SNIa were investigated as an additional site \cite{Kusakabe05}. In total, the same trend was observed as shown in Figure \ref{fig:over} for SNII. The underproduction of the Mo-Ru p-nuclei was less pronounced perhaps due to the slightly higher temperatures. Despite the fact that the total amount of p-nuclei produced in one event is higher than for SNII, the less frequent occurrence of SNIa reduces their contribution to the observed abundances \cite{Iwamoto99}. Two recent studies \cite{Kusakabe11, Travaglio11} confirm these findings, although the estimated underproduction of $^{92,94}$Mo and $^{96,98}$Ru is further decreased by an additional contribution to their abundances stemming from proton capture reactions.
Thus, a combination of both SNIa and SNII is mandatory to match the absolute observed abundances. There might be additional but small contributions from events occurring less frequently like, e.g., sub-Chandrasekar mass supernovae \cite{Goriely02} or pair-creation supernovae \cite{Arnould98}. As for SNIa, processes besides the $\gamma$-process also contribute at these more exotic sites. 

The nucleosynthesis of Ta, which has remained a puzzle over the years, is also worth mentioning. An accurate determination of the isotopic composition of Ta would enable p-process nucleosynthetic calculations to be evaluated in terms of an accurate isotope abundance for $^{180}$Ta. This nuclide is produced by both the p- and s-process and has the remarkable property of being the rarest isotope in the solar system, existing only in a long-lived isomeric state at E$_x$ = 77 keV ($t_{1/2,iso}$ $>$ $10^{15}$ $yr$) with an isotopic abundance of about 0.012\%. In reality one is measuring the isotope abundance of $^{180m}$Ta, which is a unique situation in nature. In its ground state, $^{180}$Ta decays to $^{180}$Hf and $^{180}$W with a half-life of only 8 hours. $^{180m}$Ta is the rarest isotope in nature and is, therefore, an important isotope in deciphering the origin of the p-process. Over the years many processes, such as slow- and rapid neutron capture reactions in stars and ccSN explosions, photon- and neutrino-induced reactions in ccSNe, have been proposed as the production mechanism of $^{180}$Ta. However, no consensus exists and it has been theoretically shown that $^{180}$Ta could be exclusively explained with the $\gamma$-process ($\gamma$,n) \cite{Rayet95}. The s-process alone can exclusively explain the production of $^{180}$Ta, as well, mostly via branching in $^{179}$Hf through the reactions $^{179}$Hf($\beta^{-}$)$^{179}$Ta(n,$\gamma$)$^{180}$Ta and/or $^{179}$Hf(n,$\gamma$)$^{180m}$Hf($\beta^{-}$)$^{180}$Ta \cite{Loewe03}. Furthermore, more exotic reactions such as neutrino processes, which include $^{180}$Hf($\nu_{e}$,$e$)$^{180}$Ta, have been proposed to partly explain its synthesis \cite{Heger05, Byelikov07}. 
Nevertheless, the significance of individual processes cannot be clearly determined as a result of the uncertainties in the reaction rates for $^{180}$Ta due to the unavailability of experimental data, such as the $\gamma$-ray strength function \cite{Goriely01}. An accurate determination is required to provide a better basis for p-process production calculations \cite{Laeter05}. Recently, a high-precision method has been developed to measure the isotopic ratios of extraterrestrial samples with low concentrations of Ta, but the extreme difference in isotopic abundances of a factor of more than 8000 makes the precise and accurate determination of Ta isotope ratios by mass spectrometry very challenging (see, for more details, \cite{Pfeifer17}).

\subsection{Neutron star and neutron star-black hole mergers: r-process}
\label{sec:NSBH}

This original site for the production of r-process nuclei was proposed by Tsuruta $et$ $al$ \cite{Tsuruta65} early in the development of the theory of nucleosynthesis. It relies on the fact that at high densities (typically $\rho$ $>$ $10^{10}$ g cm$^{-3}$) matter tends to be composed of nuclei lying on the neutron-rich-side of the valley of nuclear stability as a result of endothermic free-electron captures \cite{Goriely11}. Such conditions are found in the compression of the matter when a NS forms and in the merger of two NSs, making these systems promising sites of heavy r-process elements \cite{Rosswog14, Wanajo14, Rosswog17}. It was estimated that   5\% of the original mass of a NS may be ejected during tidal disruption of the NS in a NS-BH merger \cite{Lattimer76, Lattimer77}. Recent estimates for the amount of cold NS matter ejected during a NS merger range from $\sim$ $10^{-4}$ $M_{\odot}$ to $\sim$ $10^{-2}$ $M_{\odot}$ \cite{Hotokezaka13}, with velocities 0.1-0.3c. For NS-BH mergers, the ejecta can be up to $\sim$ 0.1 $M_{\odot}$, with similar velocities \cite{Kyutoku15}. Most of the dynamical ejecta originate from the contact interface between the colliding binary components, which deform into drop-like shapes prior to the merger, as seen in Figure \ref{fig:merg}. Subsequently, the shock-heated matter is expelled by quasi-radial pulsations of the remnant in a broad range of angular directions. For a 1.35-1.35 $M_{\odot}$  binary merger the ejecta in the shear interface between the stars are separated into two components, each being fed (nearly) symmetrically by material from both colliding stars. The mass ratio also influences the ejected mass, with very asymmetric binaries generating up to about twice the material as a symmetric binary of the same total mass \cite{Bauswein13}. Recent works have used detailed hydrodynamic simulations of mergers of two NSs to find a robust production of r-process nuclei with $A$ $\gtrsim$ 130 (e.g., \cite{Fern13, Bauswein13}). Based on these studies, the extremely neutron-rich ejecta are heated by $\beta$-decay during their decompression and can also be shocked to high temperatures during their dynamic ejection. Due to the very high initial density of the dynamical ejecta, heavy nuclei are already present during the nuclear statistical equilibrium phase of the expansion. The subsequent hot r-process undergoes fission cycling, thereby producing a stable abundance pattern for $A$ $\gtrsim$ 130. Important results have been obtained from recent research (e.g., \cite{Rosswog13, Perego14, Martin15, Sekiguchi16, Radice16, Wu16}) including simulations that consider both the composition of the dynamical ejecta and the neutrino wind (along the poles), where matter is ejected from the hot NS up to the point of BH formation, followed by the ejection of matter from (viscous) BH accretion disks. The main aspects of these studies can be summarized as follows: the dynamical ejecta mass depends weakly on the mass ratio and significantly on the binary asymmetry degree; outflows from BH accretion discs formed in NS mergers provide an important contribution to the r-process yields of compact binary mergers; eccentric binaries can eject orders of magnitude more mass than binaries in quasi-circular orbits and only slightly less than NS-BH mergers. 
In NS-BH mergers \cite{Korobkin12, Foucart14, Mennekens14, Kyutoku15, Roberts17}} the primary mechanism of mass ejection is the tidal force that disrupts the NS on the equatorial plane via angular momentum redistribution \cite{Hotokezaka13}. The geometry of the ejecta is thus fundamentally different from that of NS mergers, as Figure \ref{merg_NSBH} illustrates. Also, the ejecta from NS-BH mergers often cover only part of the azimuthal range \cite{Kyutoku15}.

\begin{figure}
\resizebox{0.48\textwidth}{!}{%
  \includegraphics{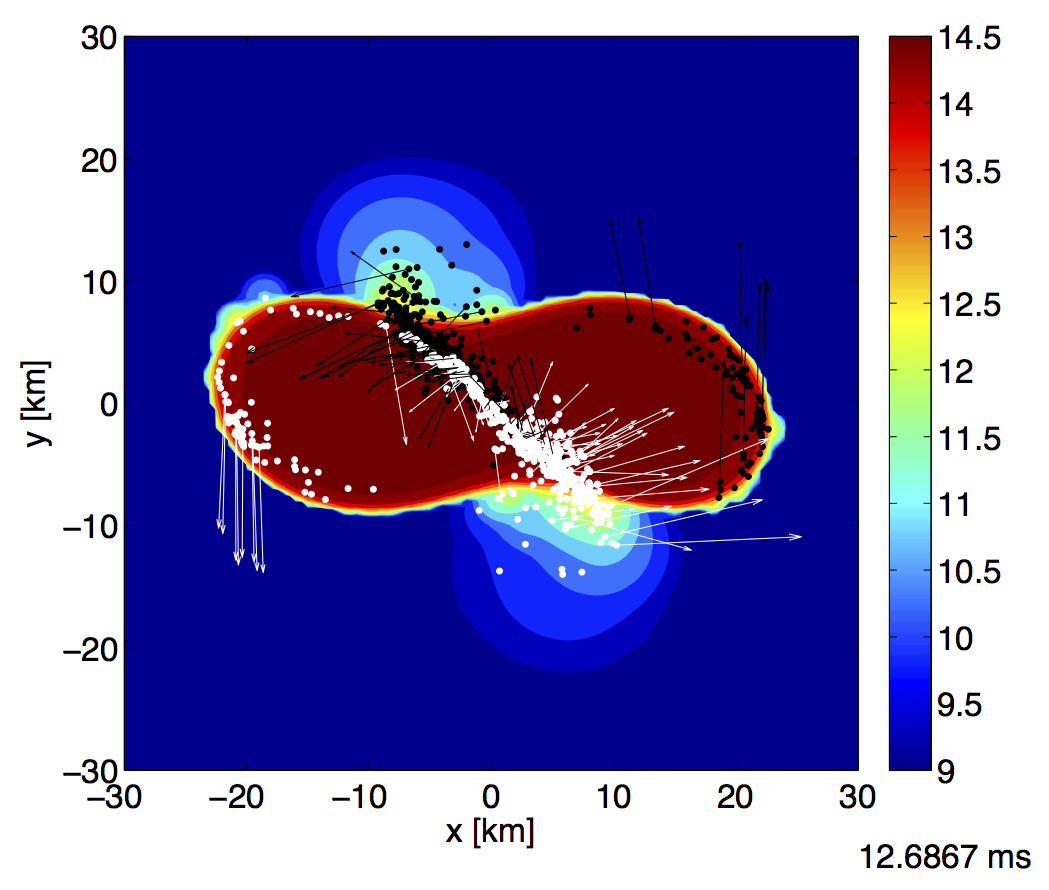}
}
\caption{Merger and mass ejection dynamics of a 1.35-1.35 $M_{\odot}$ binary neutron-star with the nuclear equation of state DD2, visualized by the color-coded conserved rest-mass density (logarithmically plotted in g/cm$^{3}$) in the equatorial plane. The dots mark particles which represent ultimately gravitationally unbound matter. Credit: \cite{Bauswein13}, reproduced with permission.}
\label{fig:merg}       
\end{figure}

One interesting aspect to be discussed relates to the ejected nucleosynthesis composition from compact object mergers. The nucleosynthesis is constrained by solar r-process abundances and by observations of low-metallicity stars. NS and NS-BH mergers seem to significantly contribute to the galactic r-process abundance pattern. However, the results obtained by different studies are conflicting or inconclusive. For instance, the overall amount of heavy r-process material in the Milky Way is consistent with the expectations of mass ejection in numerical merger simulations \cite{Radice16} with their expected rates as estimated from galactic NS mergers (e.g., \cite{Kim15}). Furthermore, recent studies obtained by Matteucci $et$ $al$ \cite{Matteucci14} point out how r-process elements originating in NS binary mergers seems to represent the most promising channel for r-process element production these days. In contrast, estimates of the impact of such double NS mergers on the galactic nucleosynthesis was questioned by detailed inhomogeneous chemical evolution studies \cite{Argast04} which are not consistent with observations at very low metallicities. The reason for the reported differences is probably due to the fact that the model proposed by Argast $et$ $al$ \cite{Argast04} does not assume instantaneous mixing in the early galactic evolutionary phases. In the study reported by Vangioni $et$ $al$ \cite{Vangioni16}, the r-process evolution using the NS scenario as the main astrophysical site is in good agreement with observations, assuming that the early evolution is dominated by mergers of binary systems with a coalescence timescale of the order of $\sim$ 100 Myr. Such mergers represent a significant fraction of all mergers according to recent estimates obtained with detailed population synthesis codes. Furthermore, several recent works \cite{Voort15, Montes16} have confirmed that the enrichment history and distribution of various r-process elements in galaxies can be accounted by NS mergers. 

A new theoretical model proposes that ccSNe  contribute first to the enrichment of heavy elements in the early galaxy.  NS mergers then follow and gradually change abundances to the solar system once \cite{Shibagaki16}. The model predicts several specific observational evidences for the time evolution of the isotopic abundance pattern. It also satisfies the universality of the observed abundance pattern between the solar system and extremely metal-poor stars in the Milky Way halo or recently discovered ultra-faint dwarf galaxies \cite{Kajino17}. Models based on particle hydrodynamics codes \cite{Hirai15} and a detailed abundance analysis of dwarf galaxies \cite{Roederer16} strongly support the argument that NS mergers are the major astrophysical site of the r-process. However, recently Bramante $et$ $al$ \cite{Bramante16} have claimed that NS mergers are unlikely to produce the r-process overabundance observed in the Reticulum II dwarf galaxy, since the total production rate of NS mergers is low, and supernova natal kicks efficiently remove binary stellar systems from the shallow gravitational well of the galaxy. A second problem that arises is that dwarf galaxies are composed of a very old stellar population \cite{Brown14}, suggesting that the chemical abundances have been frozen for $\approx$ 13 Gyr. This would require that the r-process formation  take place relatively soon after the formation of the first stars. This raises the question whether mergers could take place sufficiently rapidly so that their r-process material would be able to enrich the old stellar population. Despite this, the first direct detection of gravitational waves from a binary NS merger (GW170817) marked the true beginning of joint gravitational wave-electromagnetic multi-messenger astronomy \cite{Abbott17b} and placed stronger constraints on r-process enrichment from NS mergers. The masses ejected are broadly consistent with the estimated r-process production rate required to explain the Milky Way r-process abundances, providing the first evidence that binary NS mergers are the dominant source of heavy r-process nuclei in the galaxy \cite{Chornock17, Cowperthwaite17}. Finally according to Foucart $et$ $al$ \cite{Foucart14},  NS-BH mergers may also contribute to the enrichment of r-process elements in galaxies. According to this study, a large amount of neutron-rich, low entropy material is ejected (0.04 $M_{\odot}$-0.2 $M_{\odot}$), that will undergo robust r-process nucleosynthesis, although the ejecta is more proton-rich than the material ejected during NS binary mergers.

\begin{figure}
\resizebox{0.45\textwidth}{!}{%
  \includegraphics{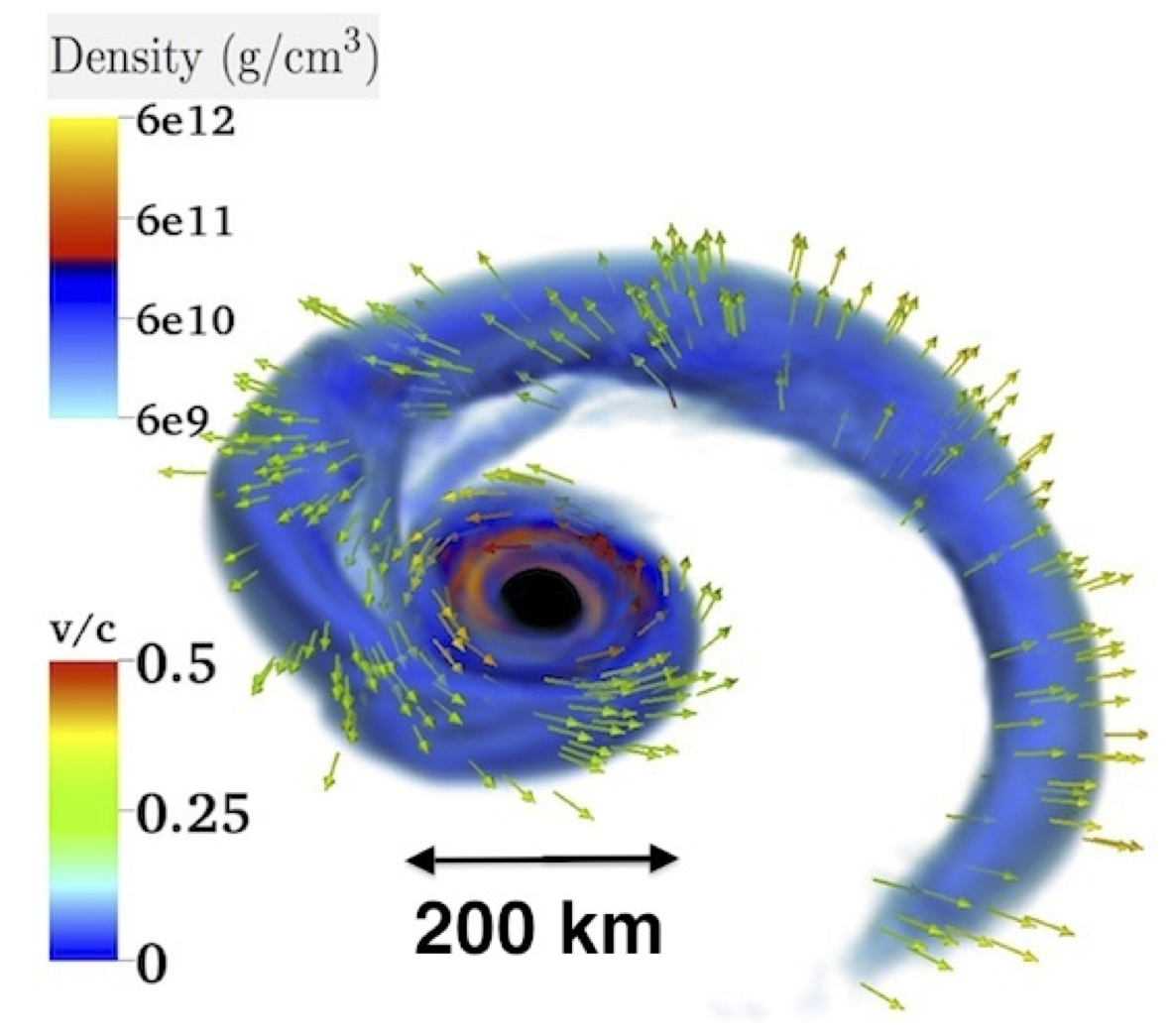}
}
\caption{Density rendering from a NS-BH merger with mass ratio 1.2/7 $M_{\odot}$ (from \cite{Foucart14}). The ejecta are confined to the equatorial plane because they are generated primarily by tidal forces.}
\label{merg_NSBH}       
\end{figure}

A reliable estimate of the NS merger rate in the galaxy is crucial in order to predict their contribution to the r-process element enrichment. Estimates of this rate are rather low because we know only few of such systems with merger times less than the age of the universe. Two of the observed binary NS systems in our Galaxy, the PSR J0737-3039 \cite{Burgay03}, and the PSR 2127+11C \cite{Prince91}, will merge in less than a few hundred Myr due to orbital decay caused by gravitational radiation emission. The total time from birth to merger is $\approx$ 8 $\times$ $10^{7}$ yr for PSR J0737-3039 and $\approx$ 3 $\times$ $10^{8}$ yr for PSR 2127+11C. Estimates for the rate of NS mergers in the galaxy range from $\sim$ $10^{-6}$ to $\sim$ 3 $\times$ $10^{-4}$ yr$^{-1}$ with the best guess being $\sim$ $10^{-5}$ yr$^{-1}$ (e.g., \cite{Kalogera04, Belczynski07}). The birth rates of NS-BH and NS binaries are comparable. Nevertheless, the fraction of NS-BH binaries having the appropriate orbital periods for merging within the age of the universe ($\sim$ $10^{10}$ yr) is uncertain due to their complicated evolution involving mass exchange \cite{Phinney91}. In any case, the total rate of NS (including NS-BH) mergers in the galaxy is perhaps $\sim$ $10^{-5}$ yr$^{-1}$, which is $\sim$ $10^{3}$ times smaller than the galactic rate of SNII \cite{Cappellaro99}. This means that each merger must eject  $\gtrsim$ $10^{-3}$ $M_{\odot}$ of r-process material if NS mergers were solely responsible for the solar r-process abundances associated with the peaks at $A$ = 130 and 195 ($\sim$ $10^{-6}$--$10^{-5}$ $M_{\odot}$ of r-process material is required from each event in the case of ccSNe) \cite{Qian03}. Alternative scenarios based on strange star - strange star mergers have also been proposed to account for the nucleosynthesis following the merger of compact objects \cite{Paulucci17}. In particular, the most prominent feature would be the total absence of lanthanides with a mass buildup populating the low-mass ($A$ $<$ 70) region. The exact composition of NSs is still under debate and quark matter represents one of the most considered possibilities \cite{Malheiro03}. New tools and developments in this field are needed since the nucleosynthesis output from NS mergers is still uncertain and the existence of multiple r-process sites cannot yet be ruled out. For a deeper study of strange quark stars, see \cite{Weber07}.

\subsubsection{The gamma-ray burst--kilonova scenario}
\label{sec:gamma}

GRBs are flashes of gamma rays associated with extremely energetic explosions that are observed in distant galaxies (as their origin is extragalactic, they are isotropically distributed in the sky). They are the brightest electromagnetic events known to occur in the universe and  last from milliseconds to several minutes. GRBs come in two varieties - long and short - depending on how long the flash of gamma rays lasts. The energy released in each explosion varies between 10$^{50}$ and 10$^{54}$ erg. In general, about one burst per day is detected. A characterizing feature of GRBs is the observation of an X-ray glow (afterglow), which is created when the high-speed jet of particles interacts with the surrounding environment and which persists for days at the GRB location. Short GRBs result from the collision of two NSs or a NS and a BH, while long GRBs are linked to ccSNe. 
As discussed before, cataclysmic events like GRBs are strongly thought to be sites of heavy elements production. More details on GRBs can be found in \cite{Piran04, Ruffini15}.

\begin{figure}
\resizebox{0.48\textwidth}{!}{%
  \includegraphics{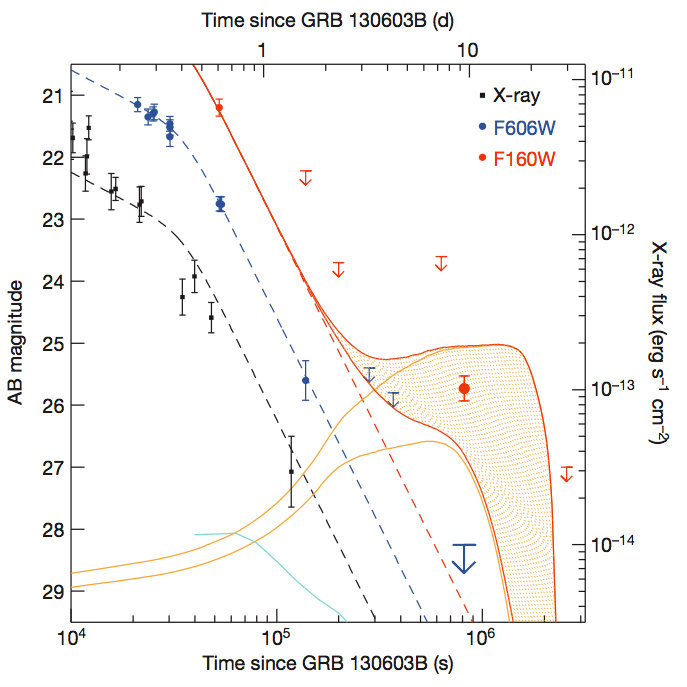}
}
\caption{Light curves of the KN seen in GRB 130603B \cite{Tanvir13}. The points represent the X-ray (black) optical (blue) and IR (red) photometry of the afterglow, along with their expected decay. The excess near-IR flux can be explained by emission powered by r-process radioactive elements produced by the ejection of neutron-rich matter during the merger of the compact objects.}
\label{fig:kilo}       
\end{figure}

In a recent work, Berger $et$ $al$ \cite{Berger13} have estimated that  the amount of Au produced and ejected during an optical/near-infrared (NIR) transient known as a Kilonova (KN), may be as large as 10 moon masses. A KN is thought to be the NIR counterpart of the merging of two compact objects in a binary system and it indicates the presence of r-process elements. It is 1,000 times brighter than a nova, but it is only 1/10th to 1/100th the brightness of an average supernova. The basic properties of KNe can be found in \cite{Metzger10}. The group studied the fading fireball from the first clear detection of a KN, which was in association with the short GRB 130603B. GRB 130603B, detected by the Swift satellite, lasted for less than two-tenths of a second. Although the gamma rays disappeared quickly, GRB 130603B also displayed an afterglow dominated by NIR light whose brightness and behavior did not match a typical afterglow. Instead, the glow behaved like it came from exotic radioactive elements. The neutron-rich material synthesized in dynamical and accretion-disk-wind ejecta during the merger can generate such heavy elements, through r-process, which then undergo radioactive decay emitting a glow that is dominated by NIR light. Calculations say that $\sim$ $10^{-2}$ $M_{\odot}$ of material was ejected by the GRB, some of which was Au and Pt. By combining the estimated Au produced by a single short GRB with the number of such explosions that have occurred over the age of the universe, all the Au in the universe might have come from GRBs. In figure \ref{fig:kilo} the interpolation of optical and NIR emissions of GRB 130603B to the F606W and F160W filters is shown. The optical afterglow decays steeply after the first $\sim$ 0.3 days and is modeled here as a smoothly broken power law (dashed blue line).  The key conclusion from this plot is that the source seen in the NIR requires an additional component above the extrapolation of the afterglow (red dashed line) \cite{Tanvir13}. This excess NIR flux corresponds to a source with absolute magnitude of $\sim$ -15.35 at $\sim$ 7 days after the burst in the rest frame. The re-brightening in the NIR afterglow is what one would expect from a KN \cite{Barnes13}. Further observational evidence of the GRB-KN connection is given in \cite{Yang15, Jin16}.

Numerical simulations show that KN scenarios can eject a small part of the original system into the interstellar medium \cite{Goriely11} and also form a centrifugally supported disk that is quickly dispersed in space with a neutron-rich wind \cite{Kasen13}. These two different ejection mechanisms are characterized by  material of differing compositions.
The outflows from the disk are likely lanthanide-free since the synthesis of heavier elements is suppressed by the high temperature \cite{Barnes13}, while the surface material is the site of an intense r-process nucleosynthesis, producing heavy elements. According to Kasen $et$ $al$ \cite{Kasen13}, the intimate relationship between KNe and the production of r-process elements makes the transient a powerful diagnostic of the physical conditions in the merger. This feature arises from the sensitivity of the optical opacity to the type of r-process composition of the ejecta: even a small fraction of lanthanides or actinides ($A$ $>$ 140) can increase the optical opacity by orders of magnitude relative to iron-group-like composition. The KN transient produces optical emission for the first day after the merger, then evolves to the NIR. The peak optical and infrared luminosities, as well as the transient duration, are increasing functions of the total ejected mass. A substantial amount of blue optical emission is generated by the lanthanide -rich ejecta at early times when the temperatures are high. The duration of this signal is $<$ 1 day \cite{Fern17}. Further calculations and atomic structure models are needed to fully establish the early time KN colors since the reliability of the predicted optical emission is affected by the uncertainties in the lanthanide  atomic data.

The fresh and revolutionary joint detection of gravitational and electromagnetic radiation from a single source, GW170817, produced by the merger of two NSs, strongly supports the connection between short GRBs and the following KNe powered by the radioactive decay of r-process species synthesized in the ejecta \cite{Arcavi17, Kasen17, Pian17, Tanvir17}. The thermal spectrum of the optical counterpart of GW170817 (e.g. \cite{Nicholl17}) is in agreement with the KN model, as compared to the power-law spectrum expected for non-thermal GRB afterglow emission. The shape of the bolometric light curve following peak is broadly consistent with the $\propto$ $t^{-1.3}$ radioactive heating rate from freshly synthesized r-process nuclei \cite{Metzger10, Cowperthwaite17}. The light curves exhibit a rapid decline in the bluest bands, an intermediate decline rate in the red optical bands, and a shallow decline in the NIR. The total mass of the red (lanthanide-bearing) ejecta was estimated to be $\approx$ 4 $\times$ $10^{-2}$ $M_{\odot}$ with a somewhat lower expansion velocity, v $\approx$ 0.1c, than the blue ejecta. The red KN emitting ejecta component dominates the total ejecta mass and thus also likely  dominates the yield of both light and heavy r-process nuclei. Assuming an r-process abundance pattern matching the solar one, one infers that $\sim$ 100-200 $M_{\oplus}$ in Au and $\sim$ 30-60 $M_{\oplus}$ in U were created within a few seconds following GW170817 \cite{Arnould07}. Future developments in this field at the intersection of nucleosynthesis, GW astronomy, and galactic chemical evolution promise to be exciting.

\subsection{Accreting neutron stars: rp-process} 

\label{sec:rp-nuc}

Nuclei close to the proton drip-line are crucial in both quiescent and explosive astrophysical scenarios. Conditions suited for the synthesis of nuclides in the range of the p-nuclei are also established by explosive scenarios such as, X-Ray Bursts (XRBs) and X-ray pulsars, which represent possible sites for the astrophysical rp-process \cite{Schatz99}. The rp-process consists of a series of rapid proton and $\alpha$-capture reactions, interspersed with $\beta^{+}$ decays, that drives the reaction path close to the proton drip-line. Nuclear properties such as masses, lifetimes, level densities and spin-parities of states for many nuclei close to the proton drip-line must be known to fully understand the rp-process. The rp-process is inhibited by $\alpha$-decay, which puts an upper limit on the endpoint at $^{105}$Te \cite{Schatz01}. XRBs occur in binary stellar systems where a compact NS accretes H- or He-rich material from a companion star \cite{Schatz98}. Type I XRBs occur when accretion rates are less than $10^{-9}$ $M_{\odot}$ per year \cite{Schatz99} and are characterized by extremely energetic ($\sim$ $10^{39}$ ergs) bursts of X-ray radiation that appear in a very regular fashion on a timescale of hours-days. The bursts themselves last for tens to hundreds of seconds and are the result of the accumulation of material on the NS surface. After a few hours, a thermonuclear runaway under the extreme temperature ($\ge$ $10^{9}$ K) and density ($\rho$ $\sim$ $10^{6}$ g cm$^{-3}$) conditions triggers an explosion that gives rise to a bright X-ray burst \cite{Parikh08}. A great difficulty in the modelling of XRBs comes from the lack of clear observational nucleosynthesis constraints. A recent review of type I XRBs can be found in \cite{Strohmayer06}. Although the large gravitational potential generated by NSs is thought to prevent the rp-process from contributing to the chemical composition of the universe, knowledge regarding the rp-process is, however, crucial in understanding energy generation in XRB scenarios. Additionally, the chemical composition of the ashes that remain on the surface of the NS as a consequence of the rp-process is critically affected by the precise path and rate of progression of the thermonuclear reactions that constitute the rp-process \cite{Schatz99}. It is thought that proton-rich T\textsubscript{z} = -1 nuclei (where T\textsubscript{z} = $\frac{1}{2}$(N - Z)) in particular play a critical role in XRB scenarios \cite{Boyd98}. For instance, a recent theoretical study by Parikh $et$ $al$ \cite{Parikh13} highlighted the radiative proton capture reactions $^{61}$Ga(p,$\gamma$)$^{62}$Ge and $^{65}$As(p,$\gamma$)$^{66}$Se as reactions that critically affect the chemical yields generated in XRBs \cite{Parikh08}. As such, detailed structure information for states above the proton threshold in the T\textsubscript{z} = -1 nuclei $^{62}$Ge and $^{66}$Se is required. Consideration of mirror nuclei indicates that level densities in the astrophysically relevant energy regions are very low, rendering statistical methods, such as Hauser-Feshbach calculations, inappropriate in these cases \cite{Woods97}. Indeed, the proton capture reaction rates may be dominated by a single resonance. 

A major puzzle to be solved in rp-process studies comes from the reaction flow through the long-lived waiting points $^{64}$Ge, $^{68}$Se, and $^{72}$Kr which are largely responsible for shaping the tail of XRBs \cite{Brown02}. Of critical importance are the proton capture Q-values of these waiting points which strongly determine to which degree proton capture can bypass the slow $\beta$-decays of these waiting points. Waiting point nuclides slow down the rp-process and strongly affect burst observables. They are characterized by long $\beta$-decay half-lives of the order of the burst duration. Low or negative proton capture Q-values may hinder further proton capture because of strong ($\gamma$,p) photodisintegration. Significant progress has been made recently on the proton capture Q-value of $^{68}$Se \cite{Santo14}. The slow $\beta$-decay of the $^{68}$Se waiting point in the astrophysical rp-process can in principle be bypassed by a sequential two proton capture. The authors concluded that the $^{68}$Se(2p,$\gamma$) reaction has at best a very small effect and $^{68}$Se is a strong waiting point in the rp-process in XRBs. This provides a robust explanation of occasionally observed long burst durations of the order of minutes. Important experimental results of rp-process reaction rates have also been made recently with the GRETINA array at NSCL \cite{Langer14}. The measurements essentially remove the uncertainty in the contribution of the $^{57}$Cu(p,$\gamma$)$^{58}$Zn reaction in XRB models and also determine the effective lifetime of $^{56}$Ni, an important waiting point in the rp-process. When a NS accretes H and He from the outer layers of its companion star, thermonuclear burning processes enable the $\alpha$$p$-process (a sequence of ($\alpha$,p) and (p,$\gamma$) reactions) as a break out mechanism from the hot $CNO$-cycle. XRB models predict ($\alpha$,p) reaction rates to significantly affect light curves of XRBs and elemental abundances in the burst ashes \cite{Cyburt16}. Theoretical reaction rates used in the modelling of the $\alpha$$p$-process need to be verified experimentally. An important case in the $\alpha$$p$-process is the $^{34}$Ar($\alpha$,p)$^{37}$K reaction which has been identified in sensitivity studies \cite{Parikh08} as an important nuclear uncertainty.  Indeed, recent R-matrix calculations \cite{Long17} for several ($\alpha$,p) reactions, including $^{34}$Ar($\alpha$,p)$^{37}$K, indicate a lower than predicted cross-section. The Jet Experiments in Nuclear Structure and Astrophysics (JENSA) gas jet target \cite{Chipps14} enables the direct measurement of previously inaccessible ($\alpha$,p) reactions with radioactive beams provided by the rare isotope re-accelerator ReA3. Preliminary results have been presented of the first direct cross-section measurement of the $^{34}$Ar($\alpha$,\,p)$^{37}$K reaction \cite{Schmidt17}.

Tremendous advancements have been obtained in mass measurements of nuclei involved in the rp-process \cite{Yan13}, allowing more accurate calculations of XRBs light curves and burst ashes. Most recently, the mass of $^{31}$Cl has been measured with the JYFLTRAP \cite{Kankainen16}. The precision of the mass-excess value of $^{31}$Cl was improved from 50 keV to 3.4 keV. The mass of $^{31}$Cl is relevant for estimating the waiting point conditions for $^{30}$S, as the $^{31}$Cl($\gamma$,p)$^{30}$S-$^{30}$S(p,$\gamma$)$^{31}$Cl equilibrium ratio  depends exponentially on the Q value i.e., on the masses of $^{31}$Cl and $^{30}$S. It has been suggested that the $^{30}$S waiting point could be a possible explanation for the double-peaked type I XRBs curves observed from several sources \cite{Fisker04}. With the new Q value, photodisintegration takes over at lower temperatures than previously, and the uncertainties related to the reaction Q value have been significantly reduced.

\section{Summary and conclusions}

In summary, the measured primordial $^{7}$Li abundance falls persistently and significantly below BBN + WMAP predictions, although recently enormous efforts have been spent to experimentally investigate resonances which would amplify $^{7}$Be destruction. However, new measurements at higher neutron energy as well as better estimations of the thermonuclear rates of the involved reactions may still be needed in order to solve the long-standing cosmological Li problem. 

There are many questions about the role of neutrinos in astrophysics and cosmology. Cosmic background neutrinos are thought to contribute to dark matter and may influence large-scale structure formation. In the next few years, new massive solar neutrino detectors will generate large amounts of precise data that should have a major impact on our understanding of how the Sun shines and how neutrinos behave. Among the most important physics objectives of the Super-Kamiokande experiment is to make a significant contribution toward the understanding of solar neutrino physics. The high counting rate of $\nu$-$e$ scattering due to solar neutrinos and the capability of measuring the recoil-electron energy spectrum constitute the main points of the Super-Kamiokande approach to the solar neutrino puzzle. It will be important to observe the abundant low-energy solar neutrinos in order to test more precisely the theory of stellar evolution. Solar neutrino experiments at low energies can provide refined measurements of the parameters that describe neutrino oscillations. A broad range of neutrino detectors with low thresholds is required to make the necessary measurements. Furthermore, future recordings, tests, and data taken over a longer period of time will allow us to probe important aspects of neutrino physics.

The production of heavy elements in ccSNe is dominantly related to the neutrino-driven wind phase, in which matter is ejected from a proto-neutron star due to neutrino-nucleon interactions. Astrophysical parameters in the neutrino-driven wind, such as entropy and electron fraction, will determine which nucleosynthesis process occurs. The studies of $\nu$-nucleosynthesis in the outer shells of a star during ccSNe need better observational constraints in order to be confirmed. It would be interesting in the future to study  $\nu$-nucleosynthesis in more detail within a full detailed hydrodynamical simulation of the explosion of the star. When a star has an s-process contribution, it would be significant to study the influence of  $\nu$-nucleosynthesis on the complete set of s-process elements, and not only for a small subset of all nuclei as it has been done in previous work. Regarding r-process, coalescence of NS binaries is one of the most plausible sites for nucleosynthesis that have been extensively studied. In NS mergers, the r-process material originates in the NS crust, and the composition of the crust and how it responds to stress caused by the merger dictates the amount of r-process material which is ejected. In the near future, the target will be to build a consistent picture of NSs and the nuclear physics that governs them, informed by gravitational waves, X-ray observations and laboratory experiments.

Further research on GRBs may help scientists to understand the history of element production in the Milky Way by providing a record of events. Short GRBs are expected to create significant quantities of neutron-rich radioactive species - gold, uranium, plutonium - whose decay should result in a faint transient in the days following the burst, the so-called kilonova. The new generation of gravitational wave detectors (Advanced-LIGO and Advanced-VIRGO) has already reached sufficient sensitivity levels to detect NS and NS-BH mergers out to distances of a few hundred Mpc. The recent simultaneous detection of the electromagnetic counterpart with gravitational waves has officially begun the era of multi-messenger astronomy and has confirmed short GRBs as sites of heavy elements production. Studying the universe with these two fundamentally different types of information will offer the possibility of a richer understanding of the astrophysical scenarios as well as of nuclear processes and nucleosynthesis.

\section*{Acknowledgments}

This work is a part of the project INCT-FNA Proc. No. 464898/2014-5. The authors would like to acknowledge FAPESP for financial support under the thematic Projects 13/26258-4, 2016/17612-7 and 2017/05660-0 and through the regular research support process 2013/17696-8. Support from the CNPq is also acknowledged. 

VL received support from CAPES through the ``Science Without Borders" project, and MSH acknowledges a Senior Visiting Professorship granted by CAPES/ITA.

Finally, the authors would like to thank the referee for the careful reading of the manuscript and valuable comments that contributed to the improvement of the paper.

 \bibliographystyle{epj}
 \bibliography{epj}
%

\end{document}